\documentclass[prd,aps,showpacs,preprintnumbers,amsmath,amssymb,nofootinbib,eqsecnum, preprintnumbers]{revtex4}  
\usepackage{graphicx}
\usepackage{epsf}
\usepackage{amsmath}
\usepackage{epstopdf}
\usepackage{bm}
\usepackage{color}
\usepackage{tabularx}
\usepackage{enumitem}
\usepackage{float}
\usepackage{array,booktabs}
\usepackage{footnote}
\usepackage{threeparttable}
\usepackage{graphicx} 
\usepackage{etoolbox}
\usepackage{hyperref}
\begin{document}

\title{%
\hfill{\normalsize\vbox{%
 }}\\
{ Mixing among lowest-lying scalar mesons and scalar glueball}}

\author{Hajar Noshad
$^{\it \bf a}$~\footnote[1]{Email:
h.Noshad@shirazu.ac.ir}}

\author{S. Mohammad Zebarjad
$^{\it \bf a}$~\footnote[2]{Email:
zebarjad@shirazu.ac.ir}}

\author{Soodeh Zarepour
$^{\it \bf b}$~\footnote[3]{Email:
szarepour@phys.usb.ac.ir}}

\affiliation{$^ {\bf \it a}$ Physics Department and Biruni Observatory,  Shiraz University, Shiraz 71454, Iran,}

\affiliation{$^ {\bf \it b}$ Department of Physics, University of Sistan and Baluchestan, Zahedan, Iran}

%
%

\date{\today}

\begin{abstract}
Scalar glueball is implemented in single nonet linear sigma model (SNLSM) which basically includes lowest lying scalar and pseudoscalar mesons. Our new version of SNLSM involves mixing among scalar matter fields and glueball field which is enforced by scale symmetry considerations and the associated anomaly. Performing iterative Monte Carlo simulations, it is
found that among the three candidates of scalar glueball, i.e., $f_0(1370)$, $f_0(1500)$ and $f_0(1710)$, only $f_0(1500)$ is predominately a glueball state with the mass prediction of $1.566 \pm 0.046$ GeV and the other two are predominately quarkonium. The  $\pi\pi$, $\pi K$ and $\pi \eta$ scatterings  are reinvestigated in the presence of scalar glueball and it is found that the overall behavior of the real part of the K-matrix unitarized $\pi \pi$ scattering amplitude is more compatible with observed data compared with the case of SNLSM without glueball.
We have also presented the predictions of the model for the masses and decay widths of the scalars obtained from the poles of the K-matrix unitarized $\pi\pi$, $\pi K$ and $\pi \eta$  scattering amplitudes. Moreover the  $\pi\pi$ scattering phase shift predicted by our model is compared with the prediction of generalized linear sigma model (GLSM) which contains two nonets of scalar mesons and two nonets of pseudoscalar mesons (a quark-antiquark nonet and a four-quark nonet). 
Despite the fact that at this stage our model lacks the second meson nonet above $1$ GeV, its prediction for the $\pi\pi$ scattering phase shift is close to the prediction of GLSM for $\sqrt{s}<1.1$ GeV and in better agreement with experimental data for $\sqrt{s}>1.1$ GeV in comparison with GLSM.
\end{abstract}


\maketitle
\section{Introduction}
\noindent
The non-perturbative behavior of QCD at low-energy can be studied  using  effective field theory approaches such as
chiral perturbation theory (ChPT) \cite{scherer,weinberg,gasser,leutwyler,ecker}, single nonet linear sigma model (SNLSM) \cite{sch168,sch2874,sch733}, generalized linear sigma model (GLSM)  \cite{f_lowenergy,f_low,f_note,f_global,f_chiral,f_electro,f_trace}, nonlinear chiral Lagrangian models \cite{black_mixing,black_2,sannino,herada,f_nl,blfa} and extended linear sigma model (eLSM) \cite{pargan,jano1,haber,gallas,par2,jano2}. The basic feature of all these effective theories  is to respect the symmetries of the underlying theory such as  chiral symmetry.
\par 
 Single nonet linear sigma model (SNLSM) is a version of linear sigma model which is formulated in terms of nine scalar fields below $1$ GeV (the light and broad  isosinglet $f_0(500)$ or sigma, the isodoublet $K^*(800)$ or kappa and the two states approximately degenerate, isosinglet $f_0(980)$ and isovector $a_0(980)$) and nine pseudoscalar fields below $1$ GeV (the isosinglet eta's $\eta(547)$ and $\eta^\prime(958)$, the isodoublet $K(496)$ and the isovector $\pi(137)$). The properties of these mesons in this model are  studied based on a quark-antiquark substructure. Single nonet linear sigma model  provides a reasonable matching among  $\pi \pi$ and $\pi K$ scattering amplitudes and experimental data up to nearly $1$ GeV\cite{127thesis_soode}. However, it cannot simultaneously describe the quark contents and the mass spectrum for some scalars.  For example, according to the substructure   $q\bar{q}$ , the  mass of $K^*[800]$ with  quark content  $u\bar{s}$ seems to be heavier than the mass of $a_0(980)$ with structure $u\bar{d}$, while we know kappa meson is lighter than $a_0(980)$. Therefore,  the light scalar mesons do not follow  the purely quark-antiquark combinations and we need to include other substructures for these mesons.
  The MIT bag model \cite{22soode2,21soode2,23soode2,24soode2} gives a solution for this puzzle by considering the light scalar mesons as four-quark states. Many other models such as $K\bar{K}$ molecule, unitarized quark model and QCD sum-rules  have been applied to understand the nature of scalar \cite{kkmolecule,6reff0,8reff0,68asrar2,69asrar2,72asrar2,73asrar2,74asrar2,khodam,81asrar2,83asrar2,91asrar2,94asrar2,111asrar2,114asrar2,115asrar2,117asrar2}.

Furthermore, while the scalar mesons above $1$ GeV  are generally expected  to be  $q\bar{q}$ states, however study their masses and decay widths shows that a small component of four-quark substructure  has to be include for these mesons. 
Therefore, mixing among quark-antiquark components and four-quark components seems to be a reasonable solution to describe the quark substructure of scalars below and above $1$ GeV \cite{black_mixing,giacosa1,giacosa2,dblack}. This mixing is the basic idea of the generalized linear sigma model (GLSM)\cite{f_global,f_chiral}.  This model  is made  of two nonets of scalar mesons and two nonets of pseudoscalar mesons (one of $q\bar{q}$ type and the other of $qq\bar{q}\bar{q}$ type) for which the mixing among two and four quark nonets is preformed (See Appendix \ref{appGLSM}).
This model can greatly improve  the results of SNLSM for decay widths, masses and quark components of mesons below $1$ GeV and also in some cases  above $1$ GeV.  
Moreover, in GLSM, the obtained scattering amplitudes for $\pi \pi$ \cite{40soode2} and $\pi K$ \cite{41soode2}  are in good agreement  with the experimental data up to $1$ GeV. However, the predictions of the model for the mentioned scattering amplitudes for the energy region above 1 GeV is far from experiment. Also the model fails to obtain acceptable decay widths and masses for states above $1$ GeV. 
\par 
On the other hand we know that glueballs, bound state of glouns with integer spins, should also be considered in the Lagrangian of the model to complete the spectrum of mesons. The glueballs have not already been included in GLSM and also in SNLSM with a specific potential and as a consequence the model cannot determine the percentages of glueball components of scalar mesons. Therefore, it is inevitable to implement  scalar glueball field in the Lagrangian of this effective field theory by enforcing the scale symmetry and also taking into account trace anomaly.
  
To avoid the complexity due to the large number of involving parameters in GLSM while adding scalar glueball, it is convincing to  consider the mixing of glueball with the scalar mesons of the same quantum numbers \cite{giacosa3,alba} first in  SNLSM. Albeit our model lacks the meson nonet above $1$ GeV except for the lightest scalar glueball and also does not consider the possibility of  multiquark/molecular states at this stage, we will see that some interesting results emerge which encourages us to further study mixing of glueball with the second nonet in the framework of GLSM. 
\par 
In this paper,  we investigate the effect of  adding the lightest scalar glueball with the quantum number $J^{PC}=0^{++}$ on the properties of scalar mesons such as their decay widths, masses and quark components in the framework of SNLSM.  In order to dig deeper and understand the effect of adding scalar glueball a bit better, we have recalculated the K-matrix unitarized amplitudes of $\pi \pi$, $\pi K$ and $\pi \eta$ scatterings in the presence of the scalar glueball and compared our numerical results with the predictions of SNLSM without glueball and also the GLSM predictions.
\par 
There are three possible candidates  for the scalar glueball of our modified version of SNLSM. These  candidates are  $f_0(1370)$ (scenario I), $f_0(1500)$  (scenario II) and $f_0(1710)$  (scenario III). At present there is no agreement in literature that among the two strongest candidates, i.e., $f_0(1500)$ and $f_0(1710)$, which one is the lightest scalar glueball. While in \cite{amsler, close1,close2,asrar1,asrar2,jano1}, $f_0(1500)$ is believed to be mostly gluonic, in \cite{lee,chen}  $f_0(1710)$ was argued to be an unmixed scalar glueball. Moreover the mass obtained from Lattice calculations for the $0^{++}$ glueball candidate is around $1600-1700$ MeV with the uncertainty of $100$ MeV. It is noteworthy that in lattice calculations the mixing of glueball field with isosinglet scalar mesons was not considered \cite{lee,bali,morningstar,chen2} . 
\par 
This paper is organized as follows: In Sec. \ref{sec2},  we give a brief review of single nonet without glueball and will present its predictions for the masses and decay widths of the scalar and pseudoscalar mesons below $1$ GeV. In Sec. \ref{sec3}, we explore the effect of adding scalar glueball to the SNLSM and present the numerical results. Finally in Sec. \ref{sec4}, the results are summarized and discussed.

\def\arraystretch{1.5}
\setlength{\tabcolsep}{12pt}

\section{Brief review of the single nonet linear sigma model}\label{sec2}
\noindent
 The general form of  the Lagrangian density of the linear SU(3) sigma model is \cite{127thesis_soode}
\begin{equation}
{\cal L} = - \frac{1}{2} {\rm Tr}
\left( \partial_\mu M \partial_\mu M^\dagger
\right)- V_0 \left( M\right)- V_{SB},
\label{LsMLag}
\end{equation}
where $M$ is the $3\times 3$ chiral field constructed from scalar $S$ and pseudoscalar  $\phi$ matrices
\begin{equation}
M=S+i\phi,
\end{equation}
with the quark-antiquark substructure of
\begin{equation}
M_{a}^{b}=\left(q_{bA}\right)^\dagger \gamma_{4} \dfrac{1+\gamma_{5}}{2} q_{aA}=\left(\bar{q_{bA}}\right)_{R}\left(q_{aA}\right)_{L},
\end{equation}
where $a$ and $A$ indicate flavor and color indices respectively and $q_{L}$ and $q_{R}$ are left and right handed quark projections. Under a chiral transformation, M transforms as
\begin{equation}
M\rightarrow U_{L}M U_{R}^\dagger.
\end{equation}

Moreover, $V_{0}$ is an arbitrary function of the independent non derivative $\rm{SU(3)_{L}}\times \rm{SU(3)_{R}}\times \rm{U(1)_{V}}$ (but not necessarily $\rm {U(1)_{A}}$) invariants formed out of $M$ and $V_{SB}$ is a symmetry breaking term. Without knowing details of $V_0$ and just from chiral symmetry considerations, i.e., using generating equations, it is possible to compute the masses of pseudoscalars and some of the scalar mesons (two-point vertices) and also the three- and four-point vertices \cite{sch2874,fariborzint}. However, in such a way, the masses of the lowest lying isoscalars, $\sigma$ and $f_0(980)$, and also their mixing angle $\theta_s$, the mass of isovector $a_0(980)$ and consequently the related three-point vertices are not perfectly predicted. Hence, in this paper we prefer to make a specific choice for  $V_0$ in order to have more predictions.\\ 
We note that there is an infinite number of chiral invariant terms to be chosen for $V_0$. A systematic way for limiting the number of terms is based on the number of underlying quark and antiquark lines in each term in
the potential. Keeping the terms with twelve or fewer quark and anti quark lines at each effective vertex, the potential is given by
\begin{eqnarray}
V_0 &=&c_2 \, {\rm Tr} (MM^{\dagger}) +
c_4^a \, {\rm Tr} (MM^{\dagger}MM^{\dagger})+c_4^b \,\Big({\rm Tr} (MM^{\dagger})\Big)^2+ c_6^a\, {\rm Tr} (MM^{\dagger}MM^{\dagger}MM^{\dagger})
\nonumber \\
      &+&c_6^b \,\Big({\rm Tr} (MM^{\dagger})\Big)^3 + c_3\left[  {\rm ln} \Big(\frac{{\rm det} M}{{\rm det}
M^{\dagger}}\Big)\right]^2.
\label{withotglue}
\end{eqnarray}
Except for the last term, all the terms are invariant under $\rm{U(1)_A}$. The explicit chiral symmetry breaking term which imposes the quark masses has the minimal form
\begin{equation}
V_{SB}=-\rm{Tr}\Big(A(M+M^\dagger)\Big)=-2(A_1 S_1^1+A_2 S_2^2+A_3 S_3^3),
\end{equation}
where $A_1$, $A_2$ and $A_3$ are proportional to the three light quark masses. The ground state should satisfy the minimum condition
\begin{equation}\label{mincon}
\left< \frac{\partial V_0}{\partial S}\right>_0 + \left< \frac{\partial V_{SB}}{\partial S}\right>_0=0,
\end{equation}
where the equilibrium values of scalar and pseudoscalar fields $S,\phi$ are respectively
\begin{equation}
\langle S_b^a \rangle_0=\delta^b_a\alpha_a, \hskip 1cm
 \langle {\phi}_b^a\rangle_0=0.
\end{equation}
\par
In the isospin invariant limit, whereas A$_1$ =A$_2\ne$ A$_3$,  $\alpha_1 = \alpha_2  \ne \alpha_3 $, there are ten unknown parameters to be determined: six coupling constants ($c_2$, $c_4^a$, $c_4^b$, $c_6^a$, $c_6^b$, $c_3$), two mass quark parameters ($A_1$, $A_3$) and two vacuum values ($\alpha_1$, $\alpha_3$). The two minimum equations reduce the number of independent parameters to eight. Except for $c_3$ which only affects the isosinglet pseudoscalars properties, other Lagrangian parameters are determined using an iterative Monte Carlo simulation. This goal may be achieved by minimizing the following $\chi_0$ function which leads to the predictions of the model for physical masses and decay widths of scalars 

\begin{eqnarray}\label{chi_0}
\chi_0(c_2,c_{4}^{a},...)&=&\varSigma_{i=1}^{4}\dfrac{\rvert q_{i}^{\rm{exp.}}-q_{i}^{\rm{theo.}}(c_2,c_{4}^{a},...)\rvert}{q_{i}^{\rm{exp.}}}\nonumber
\\
&=&\dfrac{\rvert m_{\sigma}^{\rm{exp.}}-m_{\sigma}^{\rm{theo.}}\rvert}{m_{\sigma}^{\rm{exp.}}}+\dfrac{\rvert \Gamma_{\sigma}^{\rm{exp.}}-\Gamma_{\sigma}^{\rm{theo.}}\rvert}{\Gamma_{\sigma}^{\rm{exp.}}}\nonumber
\\
&&+\dfrac{\rvert m_{f_{0}(980)}^{\rm{exp.}}-m_{f_{0}(980)}^{\rm{theo.}}\rvert}{m_{f_{0}(980)}^{\rm{exp.}}}+\dfrac{\rvert \Gamma_{f_{0}(980)}^{\rm{exp.}}-\Gamma_{f_{0}(980)}^{\rm{theo.}}\rvert}{\Gamma_{f_{0}(980)}^{\rm{exp.}}}\nonumber
\\
&&+\dfrac{\rvert m_{a_{0}(980)}^{\rm{exp.}}-m_{a_{0}(980)}^{\rm{theo.}}\rvert}{m_{a_{0}(980)}^{\rm{exp.}}}+\dfrac{\rvert \Gamma_{a_{0}(980)}^{\rm{exp.}}-\Gamma_{a_{0}(980)}^{\rm{theo.}}\rvert}{\Gamma_{a_{0}(980)}^{\rm{exp.}}}\nonumber
\\
&&+\dfrac{\rvert m_{\kappa}^{\rm{exp.}}-m_{\kappa}^{\rm{theo.}}\rvert}{m_{\kappa}^{\rm{exp.}}}+\dfrac{\rvert \Gamma_{\kappa}^{\rm{exp.}}-\Gamma_{\kappa}^{\rm{theo.}}\rvert}{\Gamma_{\kappa}^{\rm{exp.}}}
\end{eqnarray}
Where $q_{i}^{\rm{exp.}}$ represent the central values of experimental masses (decay widths) of the scalars (Table \ref{table1}) and $q_{i}^{\rm{theo.}}$ denote the predictions of the model for their physical masses (decay widths).

The remaining $c_3$ parameter which only affects the $\eta$ and $\eta'$ properties can be determined using the trace  of the isosinglet pseudoscalar $2\times 2$ square mass matrix $(M_\eta^2)$
\begin{equation}
\rm{Tr}(M_\eta^2)=\rm{Tr}(M_\eta^2)_{\rm{exp.}}
\end{equation}
\par
\par
The bare masses and decay widths of scalar mesons obtained from Lagrangian (\ref{LsMLag}) are shifted to their physical values using the K-matrix unitarization method. This method which enforces the exact unitarity of the scattering amplitude, takes into account the effects of the final state interactions in $\pi \pi$, $\pi K$ and $\pi \eta$ scatterings.  The isoscalar, isodoublet and isotriplet physical masses (widths) are determined from the poles of the  the $\pi \pi$, $\pi K$ and $\pi \eta$  K-matrix unitarized amplitudes. In Appendix \ref{appA}, this method is reviewed for $\pi \pi$, $\pi K$ and $\pi \eta$ scatterings.

\def\arraystretch{1.5}
\setlength{\tabcolsep}{12pt}

\begin{table}[!htbp]
\footnotesize
\centering
\caption{Experimental masses and decay widths of isosinglet scalars below 2 GeV, isodoublet $\kappa$,  isotriplet $a_0 (980)$ and isosinglet pseudoscalars $\eta$ and $\eta'$.}
{\begin{tabular}{@{}l|lll@{}}
\noalign{\hrule height 1pt}
\noalign{\hrule height 1pt}
&   \multicolumn{2}{c}{Experiment} \\
           & Mass (GeV)  & Width (GeV)  \\
\noalign{\hrule height 1pt}
$f_0(500) $ or $\sigma$ &   $0.400\, \rm{to}\, 0.550$ & $0.400\, \rm{to}\, 0.700$ \\
$f_0(980)$ &   $0.990 \pm 0.020$ & $0.040\, \rm{to}\, 0.100$ \\
$f_0(1370) $  &  $1.200\, \rm{to}\, 1.500$ & $0.200\, \rm{to}\, 0.500$   \\
$f_0(1500) $  &  $1.505 \pm 0.006$ & $0.109 \pm 0.007$   \\
$f_0(1710) $ &   $1.720 \pm 0.006$ & $0.135 \pm 0.008$ \\
$K_0^*(800) $ or $\kappa$ &  $0.682 \pm 0.029 $ & $0.547 \pm 0.024$    \\
$a_0(980)$  & $0.980 \pm 0.020$ & $0.050 \, \rm{to}\, 0.100$  \\
$\eta$ & $ 0.547862 \pm 17\times 10^{-6}$ & $(0.131 \pm 0.00005)\times 10^{-6} $ \\
$\eta'$ & $ 0.95778 \pm 6\times 10^{-5}$& $0.000197 \pm 9\times 10^{-6} $\\
\noalign{\hrule height 1pt}
\noalign{\hrule height 1pt}
\end{tabular}
\label{table1}}
\end{table}

The typical results of the Monte Carlo minimization for the parameters of the model are given in Table \ref{para1} and for the physical masses and decay widths of the scalar and pseudoscalar  mesons are given in Tables \ref{table2} and \ref{table3}. It should be mentioned that since after running the code with different sets of random numbers, we never get $\chi_0$'s less than  $(\chi_0)_{\rm{exp.}} =1.414$, therefore we have extended the acceptable bound beyond the experimental value, for this case to $\chi_0< 3 (\chi_0)_{\rm{exp.}}$. Reported masses and decay widths in Tables \ref{table2} and \ref{table3} denote averages and standard deviations over this extended bound, i.e., the sets of masses and decay widths  for which $\chi_0< 3 (\chi_0)_{\rm{exp.}}$.  It can be seen that while the predictions of the SNLSM without including glueball in the Lagrangian, for sigma mass and decay width and also for kappa mass overlap with the experimental range, other predictions are not too close to experimental data. For pseudoscalars the predicted masses and decay constants are in the experimental range or very close to it (Table \ref{table3}). It should be emphasized that in the present order of the potential, the percentage of s-quark component of $f_0(980)$ is nearly $ 100\%$ (Table \ref{qc}) and therefore its coupling to pions is weak, i.e., $\gamma_{f_0 \pi \pi}\sim 0$, and since  $\pi \pi$ is the dominant decay channel of $f_0(980)$, the prediction of the model for decay width of $f_0(980)$ is very close to zero (Table \ref{table2}).

\begin{table}[!htbp]
\footnotesize
\centering
\caption{Typical predicted Lagrangian parameters: $c_2,\,c_4^a,\,c_4^b,\,c_6^a,\,c_6^b,\,c_3,\,A_1,\,A_3$ and vacuum parameters: $\alpha_1,\,\alpha_3$ for SNLSM without glueball for $\chi_0=2.4$.}
\label{para1}
	      	\begin{tabular}{@{}ccccc@{}}
				\noalign{\hrule height 1pt}
				\noalign{\hrule height 1pt}
			   $c_2\,{\rm (GeV}^2)$& $c_4^a$ & $c_4^b$ &$c_6^a \,{\rm (GeV}^{-2})$& $c_6^b\,{\rm (GeV} ^{-2})$  \\
			    \noalign{\hrule height 1pt}
				$-1.72\times 10^{-1}$ & $21.0$ & $4.66\times 10^{-2}$ & $7.76\times 10^{-2}$ & $4.67\times 10^{-2}$ \\
	            \noalign{\hrule height 1pt}\\
				 $c_3\,{\rm (GeV} ^4)$&$A_1 \,{\rm (GeV}^3)$ &$A_3 \,{\rm (GeV}^3)$ & $\alpha_1 \,{\rm (GeV})$& $\alpha_3 \,{\rm (GeV})$\\
				 \noalign{\hrule height 1pt}
				  $-1.73\times 10^{-4}$&$6.15\times 10^{-4}$ & $1.84 \times 10^{-2}$& $6.55\times 10^{-2}$& $9.35\times 10^{-2}$ \\			 
				\noalign{\hrule height 1pt}
				\noalign{\hrule height 1pt}
			\end{tabular}
\end{table}

\begin{figure}[!htbp]
		\begin{center}
			\epsfxsize = 1 cm
			\includegraphics[scale=0.5]{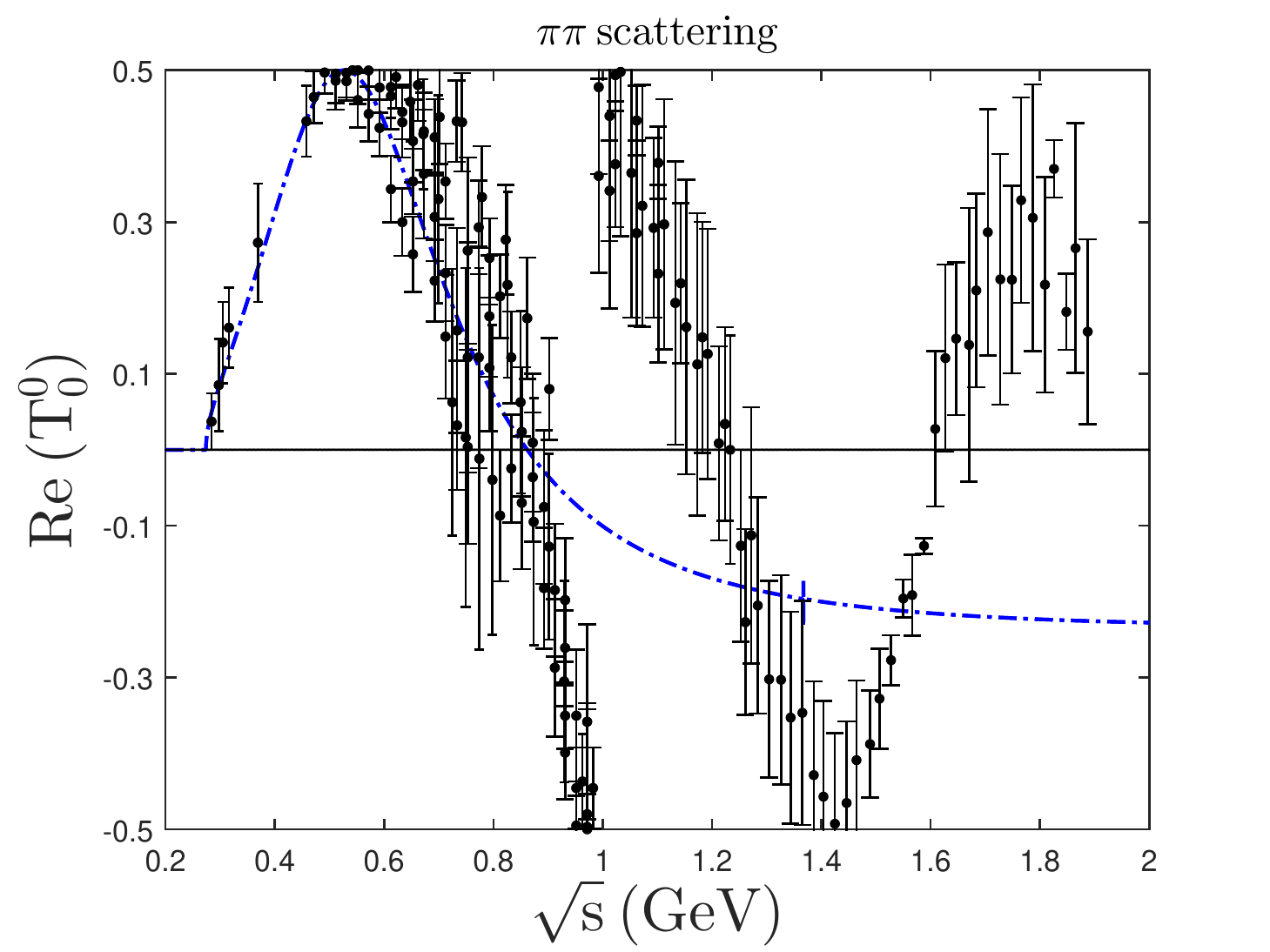}
			\hskip 0.2 cm
			\epsfxsize = 1 cm
			\includegraphics[scale=0.5]{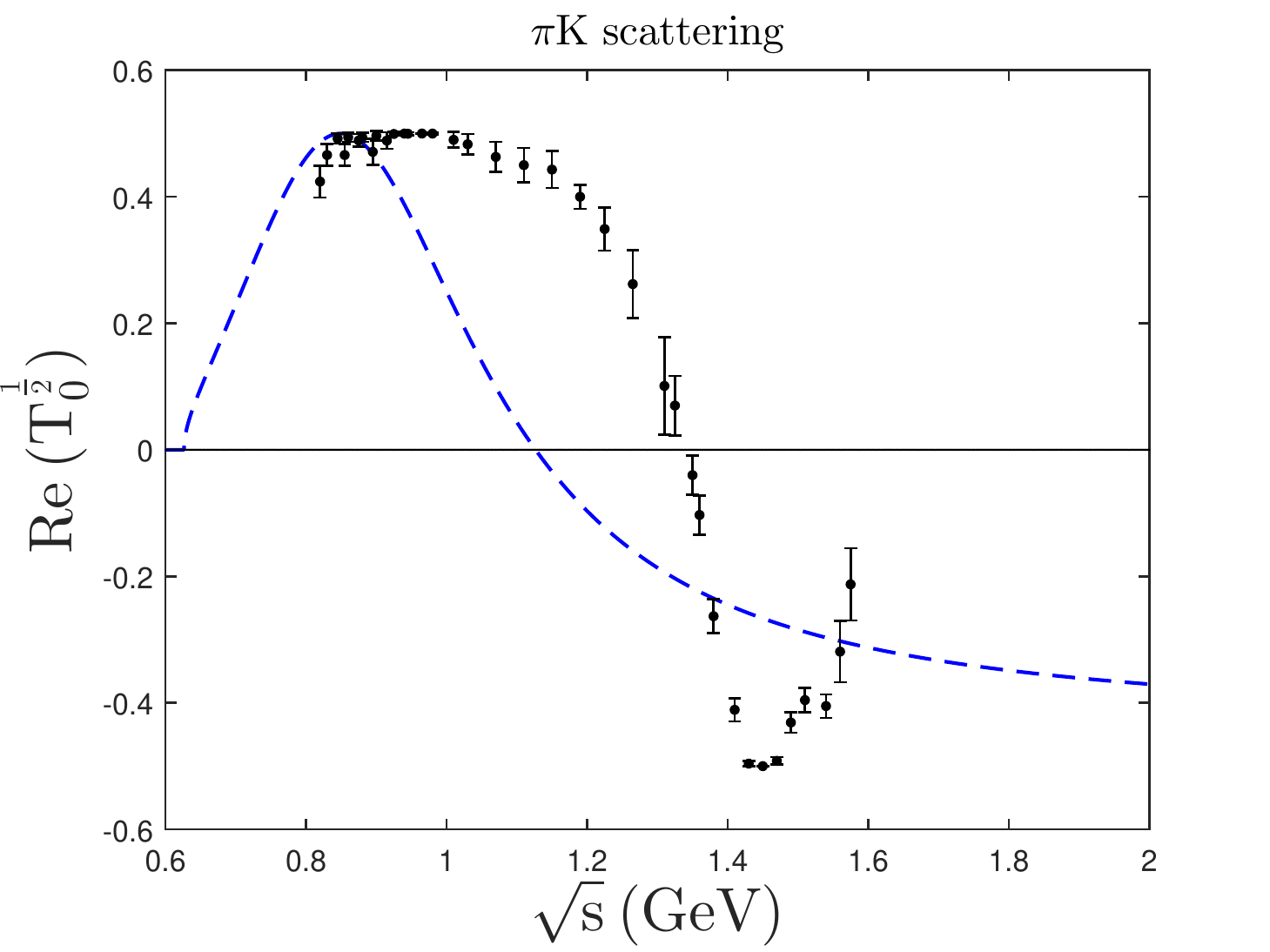}
			\vskip 0.2cm
			\epsfxsize = 1 cm
			\includegraphics[scale=0.5]{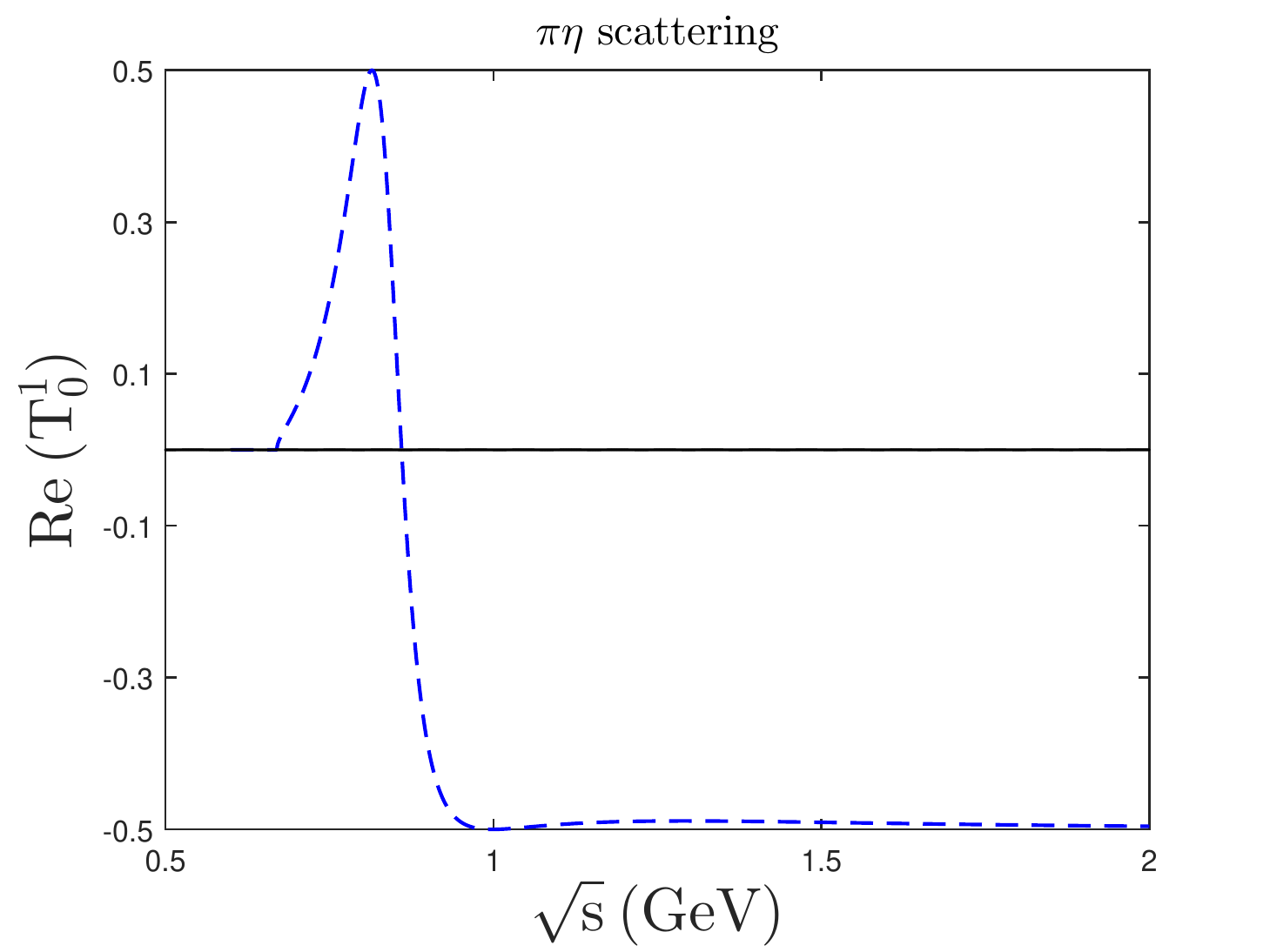}
			\caption{Predictions of SNLSM without scalar glueball for the real part of the K-matrix unitarized  $\pi \pi$, $\pi K$ and $\pi \eta$ scattering amplitudes. Up to about $900$ MeV the predictions agree with experimental data for $\pi\pi$ and $\pi K$ scatterings. There is no data for $\pi\eta$ scattering amplitude. }
			\label{plotampwg}
		\end{center}
	\end{figure}

The model predictions for the real parts of the K-matrix unitarized $\pi \pi$, $\pi K$ and $\pi \eta$   scattering amplitudes are given in  Fig. \ref{plotampwg}. The predictions agree with data up to about $900$ MeV for $\pi \pi$ and $\pi K$ scatterings while due to the lack of experimental data for the $\pi \eta$ scattering, it is not clear if the predictions are acceptable or not. For above $1$ GeV, the model lacks any structure and flattens to a constant value for all these three scattering amplitudes.

	\begin{table}[!htbp]
		\footnotesize
		\centering
		\caption{Predictions of SNLSM without including scalar glueball for masses and decay widths of scalars below 1 GeV. Note that since $\gamma_{f_0\pi \pi}$ is close to zero, the decay width of $f_0(980)$ is near to zero.}
		
		{\begin{threeparttable}
				{\begin{tabular}{@{}c|cc@{}}
						\noalign{\hrule height 1pt}
						\noalign{\hrule height 1pt}
						& Width (GeV) & Mass (GeV) \\
						\noalign{\hrule height 1pt}
						$\sigma $  &  $0.562 \pm  0.022 $& $0.454 \pm 0.001 $   \\
						$f_0(980)$ &$ 2.310\times 10^{-6}\pm  5.505\times 10^{-7} $ &  $1.362 \pm 0.229$  \\
						$\kappa$ &  $ 0.524 \pm  0.020$ &   $0.796 \pm 0.007$ \\
						$a_0(980) $ & $0.150 \pm 0.028$  &   $0.867 \pm 0.017 $  \\
						\noalign{\hrule height 1pt}
						\noalign{\hrule height 1pt}
					\end{tabular}\label{table2}}
			\end{threeparttable}}
		\end{table}

	\begin{table}[!htbp]
		\footnotesize
		\centering
		\caption{{Predictions of SNLSM without including scalar glueball for masses and decay constants of pseudoscalars below 1 GeV.}}
		{\begin{tabular}{@{}ll|ll@{}}
				\noalign{\hrule height 1pt}
				\noalign{\hrule height 1pt}
				& Mass (GeV) &  Decay constant (GeV)  \\
				\noalign{\hrule height 1pt}
				$\pi $ &   $0.137 \pm 2.8 \times 10^{-5}$ &  $0.131 \pm 4.94\times 10^{-5}$ \\
				$K$ &   $0.486 \pm 0.007$ & $0.156 \pm 0.0018 $ \\
				$\eta $  &  $0.528 \pm  0.007$ & \quad  \\
				$\eta \prime $  &  $0.968 \pm 0.004$ & \quad \\
				\noalign{\hrule height 1pt}
				\noalign{\hrule height 1pt}
			\end{tabular}\label{table3}}
	\end{table}		

	\begin{table}[!htbp]
		\footnotesize
		\centering
		\caption{The estimate of SNLSM without glueball for the percentages of strange and non-strange quark components of isoscalars.}
		{\begin{tabular}{@{}l|lll@{}}
				\noalign{\hrule height 1pt}
				\noalign{\hrule height 1pt}
				& $\frac{u\bar{u}+d \bar{d}}{\sqrt{2}}$ & $s\bar{s}$ \\
				\noalign{\hrule height 1pt}
				$f_0(500) $ &   $99.999\pm  4.311\times 10^{-5}$ & $8.499 \times 10^{-4}\pm 4.311\times 10^{-5}$ \\
				$f_0(980)$ &   $8.499  \times 10^{-4}\pm  4.311\times 10^{-5}$ & $ 99.999 \pm 4.311\times 10^{-5}$  \\
				\noalign{\hrule height 1pt}
				\noalign{\hrule height 1pt}
			\end{tabular}\label{qc}}
	\end{table}

\section{Including scalar glueball in single nonet linear sigma model}\label{sec3}
\noindent
While the classical Yang-Mills theory with massless quarks is invariant under the scale transformation $x^{\mu}\rightarrow\lambda^{-1} x ^{\mu}$, this symmetry is broken at the  quantum level (due to the fact that the  running coupling constant $g(\mu)$ depends on the energy scale $\mu$). As a consequence, the divergence of the dilatation current does not vanish and it equals to the trace of the energy-momentum tensor 

\begin{equation}\label{trace energymomemnum}
\partial_\mu D^\mu=\theta_\mu^\mu=-\dfrac{\beta(g)}{4g}G_{\mu\nu}^{a}G_{\mu\nu a},
\end{equation}
where $\theta_\mu^\mu$ is the trace of energy-momentum tensor, $G_{\mu\nu }^{a}$ denotes the Yang-Mills field strength tensor and $\beta(g)$ is the $\beta$-function given by $\beta(g)=\partial g/\partial \ln \mu=\dfrac{(11-\frac{2N_f}{3})g^3}{16\pi^2}+...$,  where $N_f$ is the number of quark flavors.  
 
 In addition, the non-vanishing expectation value of the trace anomaly is proportional to the gluon condensate
\begin{equation}\label{scaleanomaly}
\langle\theta_{\mu}^{\mu}\rangle=\dfrac{11-\frac{2N_f}{3} }{16}\left\langle  \dfrac{\alpha_s}{\pi} G_{\mu\nu}G^{\mu\nu}\right\rangle =\dfrac{11-\frac{2N_f}{3} }{16}C^4,
\end{equation}
where $\alpha_s=\dfrac{g^2}{4\pi}$ is the strong fine-structure constant and the numerical values of $C$  from QCD sum-rules and lattice QCD have been obtained  $0.3$ GeV \cite{reinders,broad,shifman} and $0.6$ GeV \cite{kripf,xue,di,ilgen,campo,gia},  respectively .

To mimic Eq. (\ref{trace energymomemnum}) in our effective theory, we need to implement a scalar glueball ($J^{PC}=0^{++}$)  $h$ field with scale dimension $1$ 
in the  Lagrangian  (\ref{withotglue}) which satisfies 
\begin{equation}\label{H}
	\theta_\mu^\mu=\dfrac{1}{\gamma^4}h^4,
\end{equation}
where $ \gamma $ is a dimensionless constant.

\par
To achieve this goal, let us consider  a potential $V_h$ constructed out of two  sets of real scalar fields $\eta_{a}$ and $\xi_{a}$, with the mass dimensions of 1 and 4 respectively. It has been shown that for this potential the trace of the energy-momentum tensor ($\theta_\mu^\mu$) reads \cite{Effective Lagrangian}
\begin{equation}\label{tetamumu}
	\theta_\mu^\mu=\Sigma_{a}(\eta_{a}\frac{\partial V_h}{\partial \eta_{a}}+4\xi_{a}\frac{\partial V_h}{\partial \xi_{a}})-4V_h,
\end{equation}
which vanishes when $V_h$ is scale invariant. 
We can  now choose   fields $\eta_a$ as   of $M$ and $M^\dagger$ and the fields $\xi_a$ as glueball $h$. Therefore, Eq. (\ref{tetamumu}) reads
\begin{equation}\label{tetamumu2}
	\theta_\mu^\mu={\rm Tr}\Big(M\frac{\partial V_h}{\partial M}+M^{\dagger}\frac{\partial V_h}{\partial M^{\dagger}}\Big)+h \frac{\partial V_h}{\partial h} -4V_h.
\end{equation}
Now, we have to look for $V_h$ in terms of $h$, $M$ and $M^\dagger$ in such a way that the obtained $\theta_\mu^\mu$ satisfies Eq. (\ref{H}). This aim is achieved by considering the terms of the form $h^4 \Sigma_m (c_m/m) {\rm ln}(R_m/\Lambda^m)$ with the constraint of $\Sigma_m c_m=1$, where $m$ is the scale dimension of $R_m$ \cite{Effective Lagrangian}. $R_m$ is an arbitrary function of $H$ and independent invariants made from $M$ and $M^\dagger$ (such as ${\rm det}(MM^\dagger)$). Following this argument, one can check with the help of Eq. (\ref{tetamumu2}), that the following potential with the constraint of $c+c'=1$, where $c$ and $c'$ are real coefficients, satisfies Eq. (\ref{H})  

\begin{eqnarray}\label{gluelag}
{\cal L}_{h} &=&- \frac{1}{2}  \partial_\mu h \partial^\mu h- V_h\nonumber\\
&=&- \frac{1}{2}  \partial_\mu h \partial^\mu h-\dfrac{c}{4\gamma^4} {\rm ln}\Big(\dfrac{h^4}{\gamma^4\Lambda^4}\Big)h^4
-\frac{c\prime}{6\gamma^4}{\rm ln}\Big(\dfrac{{\rm det}(MM^{\dagger})}{\Lambda^6}\Big)h^4,
\end{eqnarray}
where $\Lambda $ is a scale parameter with dimensions of mass and can be identified as a particular kind of QCD scale parameter.
\par
Putting Eqs. (\ref{LsMLag}) and (\ref{gluelag}) together,  the effective Lagrangian involving matter and glueball can be written as
\begin{eqnarray}
	{\cal L}=
	&-& \frac{1}{2} {\rm Tr}\left( \partial_\mu M \partial^\mu M^\dagger
	\right)- \frac{1}{2}  \partial_\mu h \partial^\mu h
	-\dfrac{c_2}{\gamma^2} {\rm Tr} (MM^{\dagger})h^2
	- c_4^a {\rm Tr} (MM^{\dagger}MM^{\dagger})\nonumber\\
	&-&c_4^b \,\Big({\rm Tr} (MM^{\dagger})\Big)^2
		-c_6^a \gamma^2 \dfrac{{\rm Tr} (MM^{\dagger}MM^{\dagger}MM^{\dagger})}{h^2}
	-c_6^b \gamma^2 \dfrac{{\Big(\rm Tr} (MM^{\dagger})\Big)^3}{h^2} 
\nonumber\\
	&-&\dfrac{c_3}{\gamma^4}\Big[ {\rm ln} \Big(\frac{{\rm det} M}{{\rm det}M^{\dagger}}\Big)\Big]^2h^4
	-\dfrac{c}{4\gamma^4} {\rm ln}\Big(\dfrac{h^4}{\gamma^4\Lambda^4}\Big)h^4
	-\frac{c\prime}{6\gamma^4}{\rm ln}\Big(\dfrac{{\rm det}(MM^{\dagger})}{\Lambda^6}\Big)h^4-2 {\rm Tr}(AS).
	\label{lagtot}
\end{eqnarray}
It is worth mentioning that the terms with coefficients $c_2$, $c_6^a$ and $c_6^b$ of the Lagrangian are modified due to scale invariance, i.e.,  the trace of the energy-momentum tensor $\theta_{\mu}^{\mu}$ for these terms must equal to zero. 
\par
 
Expanding the potential around the minimum of the dilaton field, $h=h+h_0$, and setting $\left\langle \partial V/\partial h\right\rangle =0$, $\Lambda$ is obtained as
\begin{equation}\label{lamda}
\Lambda={\rm Exp}\Big[\frac{c}{4} +\frac{\gamma^2}{2 h_0^6}\Big(c_2 h_0^4(2\alpha_1^2+\alpha_3^2)-c_6^a\gamma^4(2\alpha_{1}^6+\alpha_3^6)-c_{6}^b (2\alpha_1^2+\alpha_3^2)^3\Big)\Big]h_{0}^{c} \gamma^{-c}\alpha_{1}^{\frac{2(1-c)}{3}}\alpha_3^{\frac{(1-c)}{3}}.
\end{equation}
 Substituting $\Lambda$ in Eq. (\ref{lagtot}), we finally obtain

\begin{eqnarray}
{\cal L}=
&-& \frac{1}{2} {\rm Tr}\left( \partial_\mu M \partial^\mu M^\dagger
\right)- \frac{1}{2}  \partial_\mu h \partial^\mu h
-\dfrac{c_{2}}{\gamma^{2}}{\rm Tr}(MM^{\dagger})h^{2}- c_{4}^a{\rm Tr}(MM^{\dagger}MM^{\dagger})\nonumber\\
&-&c_{4}^b{\Big(\rm Tr}(MM^{\dagger})\Big)^{2}-c_{6}^a\gamma^{2}\frac{{\rm Tr}(MM^{\dagger}MM^{\dagger}MM^{\dagger})}
{h^{2}}-c_{6}^b\gamma^{2}\frac{{\Big(\rm Tr}(MM^{\dagger})\Big)^3}{h^{2}}\\\nonumber
&-&\Big[c_{6}^b\gamma^{2}\dfrac{(2\alpha_{1}^{2}+\alpha_{3}^{2})^3}{2h_{0}^{6}}+c_{6}^a\gamma^{2}\dfrac{(2\alpha_{1}^{6}
	+\alpha_{3}^{6})}{2h_{0}^{6}}-\dfrac{c_{2}}{\gamma^{2}}\dfrac{(2\alpha_{1}^{2}
	+\alpha_{3}^{2})}{2h_{0}^{2}}\Big]h^{4}-2 {\rm Tr}(AS)\\\nonumber
&-&\dfrac{(1-c\prime)}{\gamma^{4}}
\Big(\dfrac{-1}{4}+{\rm ln}(\dfrac{h}{h_{0}})\Big)h^4-\frac{c\prime}{6\gamma^4} \ln\Big(\frac{\det(MM^\dagger)}{\alpha_1^4\alpha_3^2}\Big)
h^{4}-\dfrac{c_{3}}{\gamma^4}\Big[{\rm ln}\Big(\dfrac{{\rm det}M}{{\rm det}M^{\dagger}}\Big)\Big]^2h^4.\\\nonumber
\end{eqnarray}

\begin{figure}[!htbp]
	\begin{center}
		\vskip 0.5 cm
		\epsfxsize = 5 cm
		\includegraphics[height=6 cm]{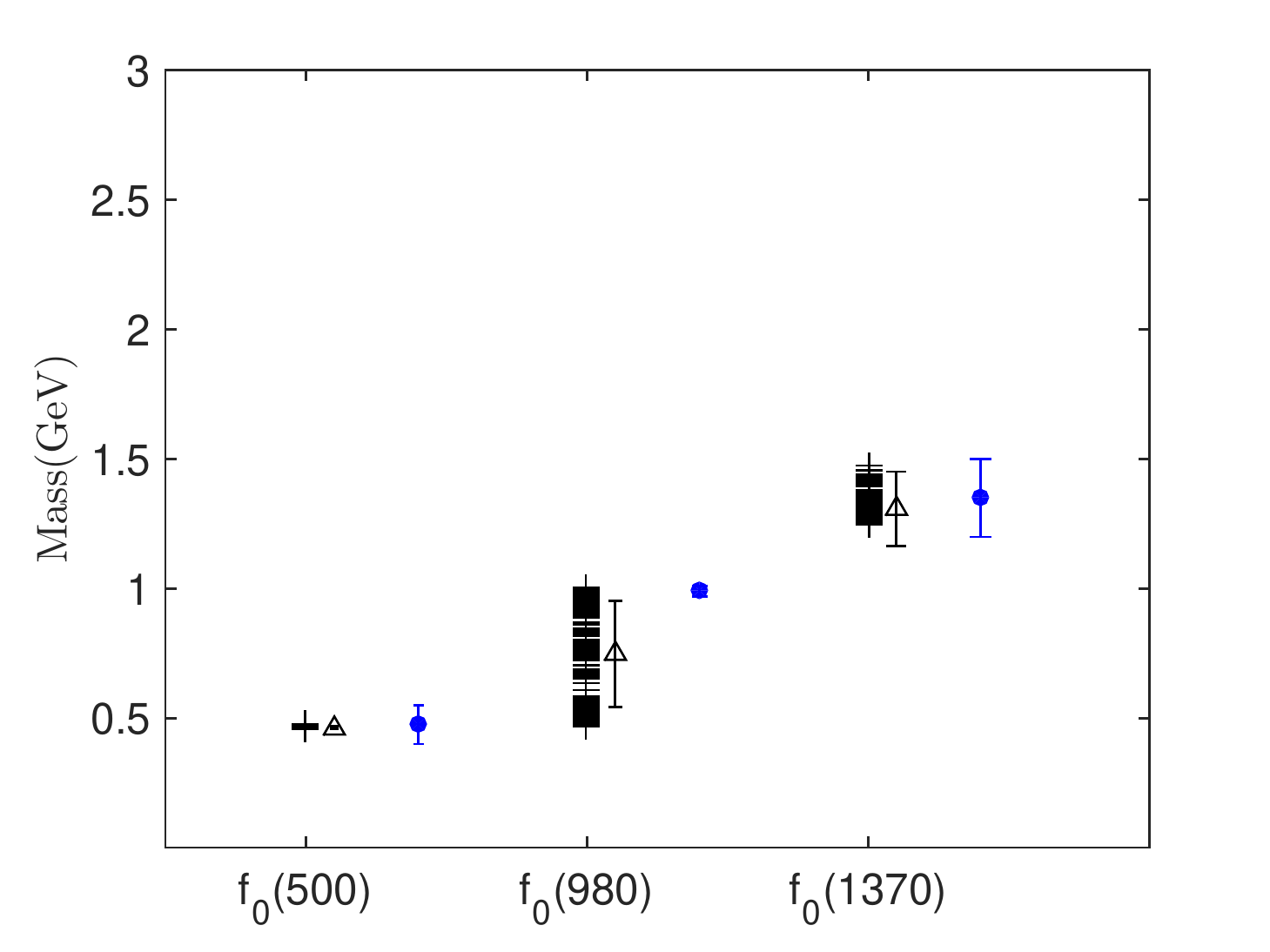}
		\hskip 1cm
		\epsfxsize = 5 cm
		\includegraphics[height=6 cm]{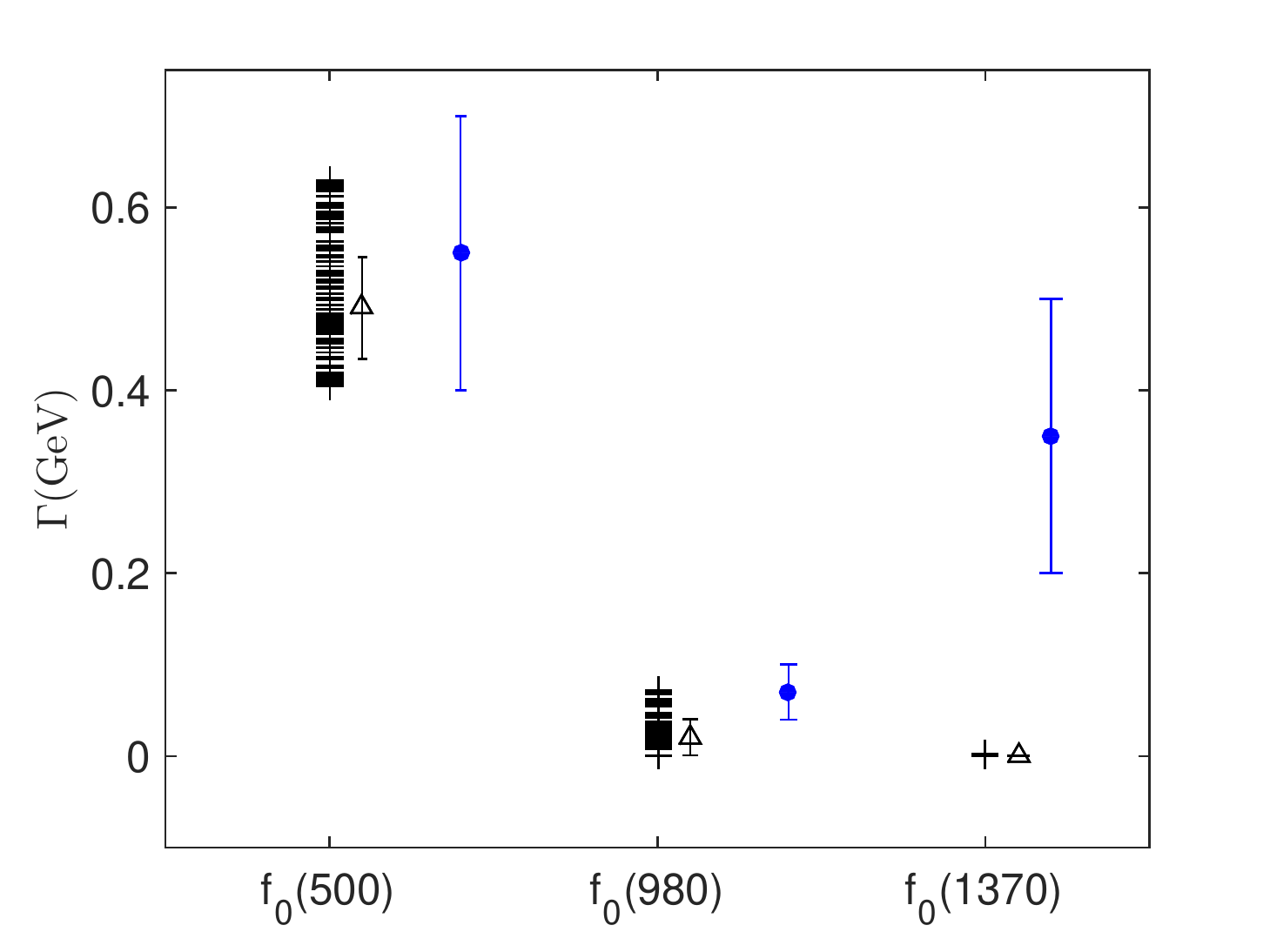}
		\caption{Masses and decay widths of isosinglet scalar mesons obtained from the Monte Carlo simulation for scenario I (pluses) are compared with their experimental values in Table \ref{table1} (solid circles with error bars). In order to compare predictions with the experimental data easier, also the average values (triangles) and  standard deviations around the averages (error bars) are depicted.  The masses and decay widths of $f_0(500)$ and $f_0(980)$ are in the experimental range but the predicted decay width of $f_0(1370)$ is too small compared with the experimental data. \label{fs1}} 
	\end{center}
\end{figure}

\begin{figure}[!htbp]
	\begin{center}
		\vskip 0.5 cm
		\epsfxsize = 5 cm
		\includegraphics[height=6 cm]{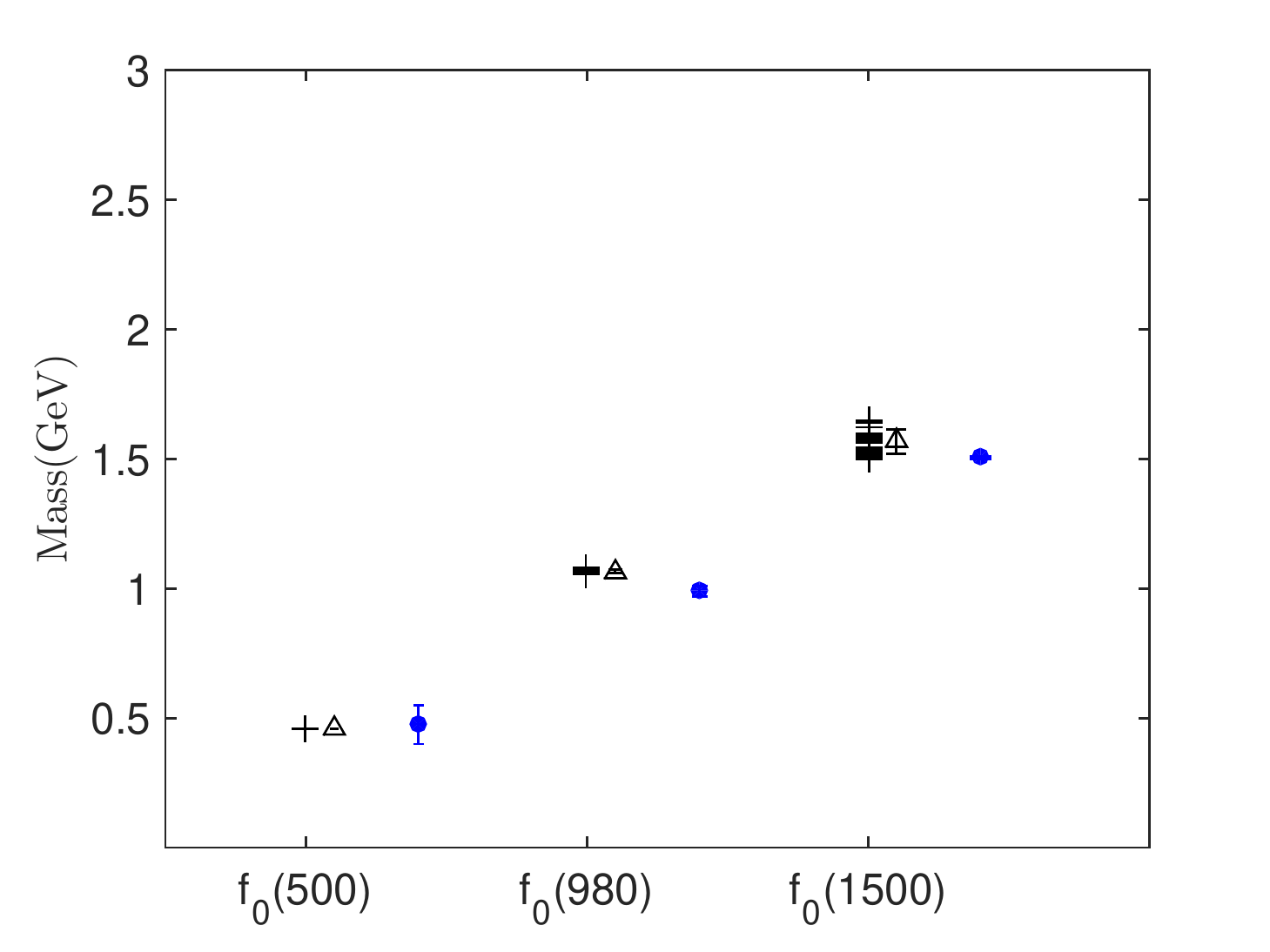}
		\hskip 1cm
		\epsfxsize = 5 cm
		\includegraphics[height=6 cm]{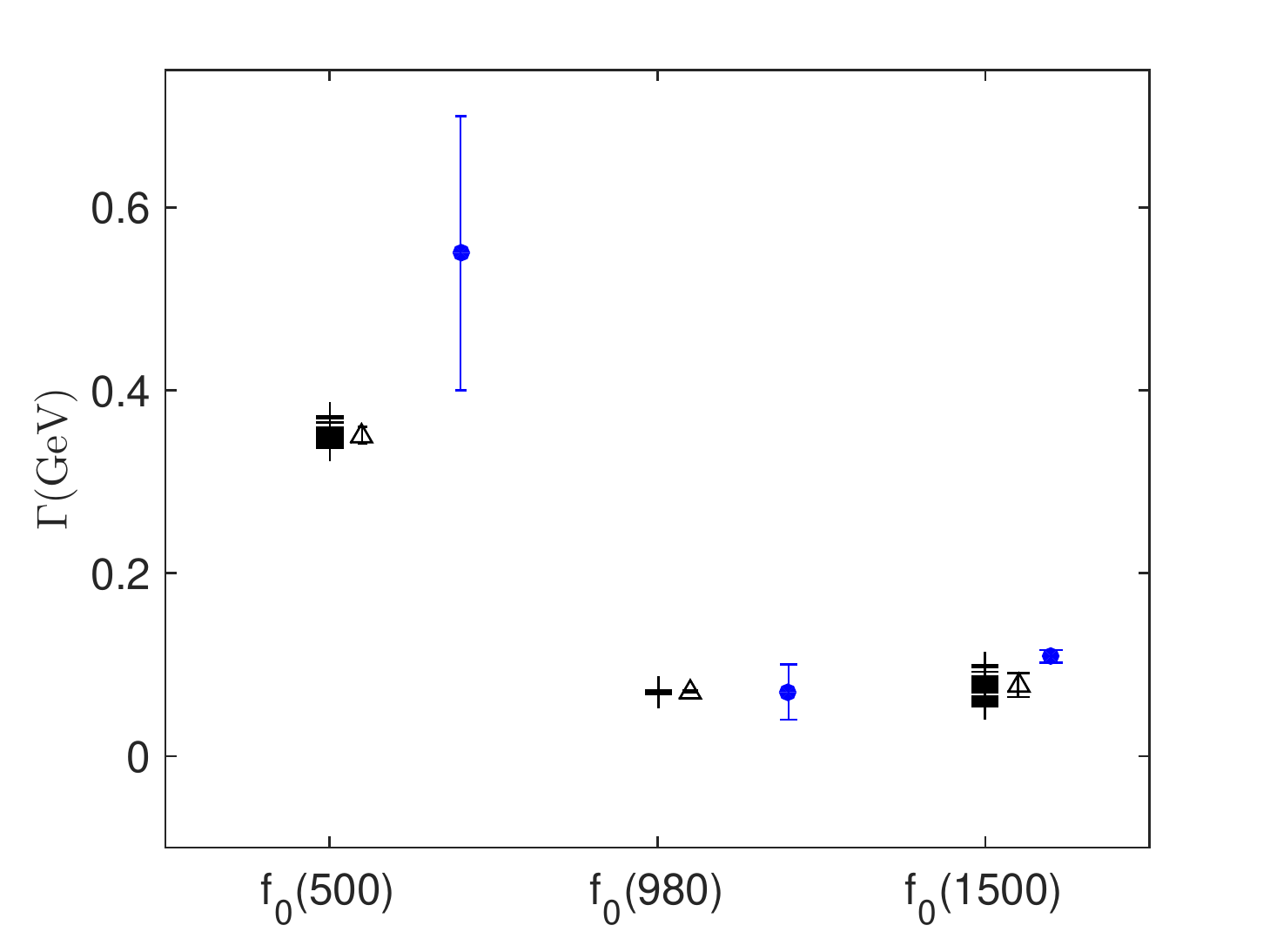}
		\caption{Masses and decay widths of isosinglet scalar mesons obtained from the Monte Carlo simulation for scenario II (pluses) are compared with their experimental values in Table \ref{table1} (solid circles with error bars). In order to compare predictions with the experimental data easier, also the average values (triangles) and  standard deviations around the averages (error bars) are depicted. The predicted masses and decay widths of all the three isoscalars are in the experimental range.}\label{fs2}
	\end{center}
\end{figure}

\begin{figure}[!htbp]
	\begin{center}
		\vskip 0.5 cm
		\epsfxsize = 5 cm
		\includegraphics[height=6 cm]{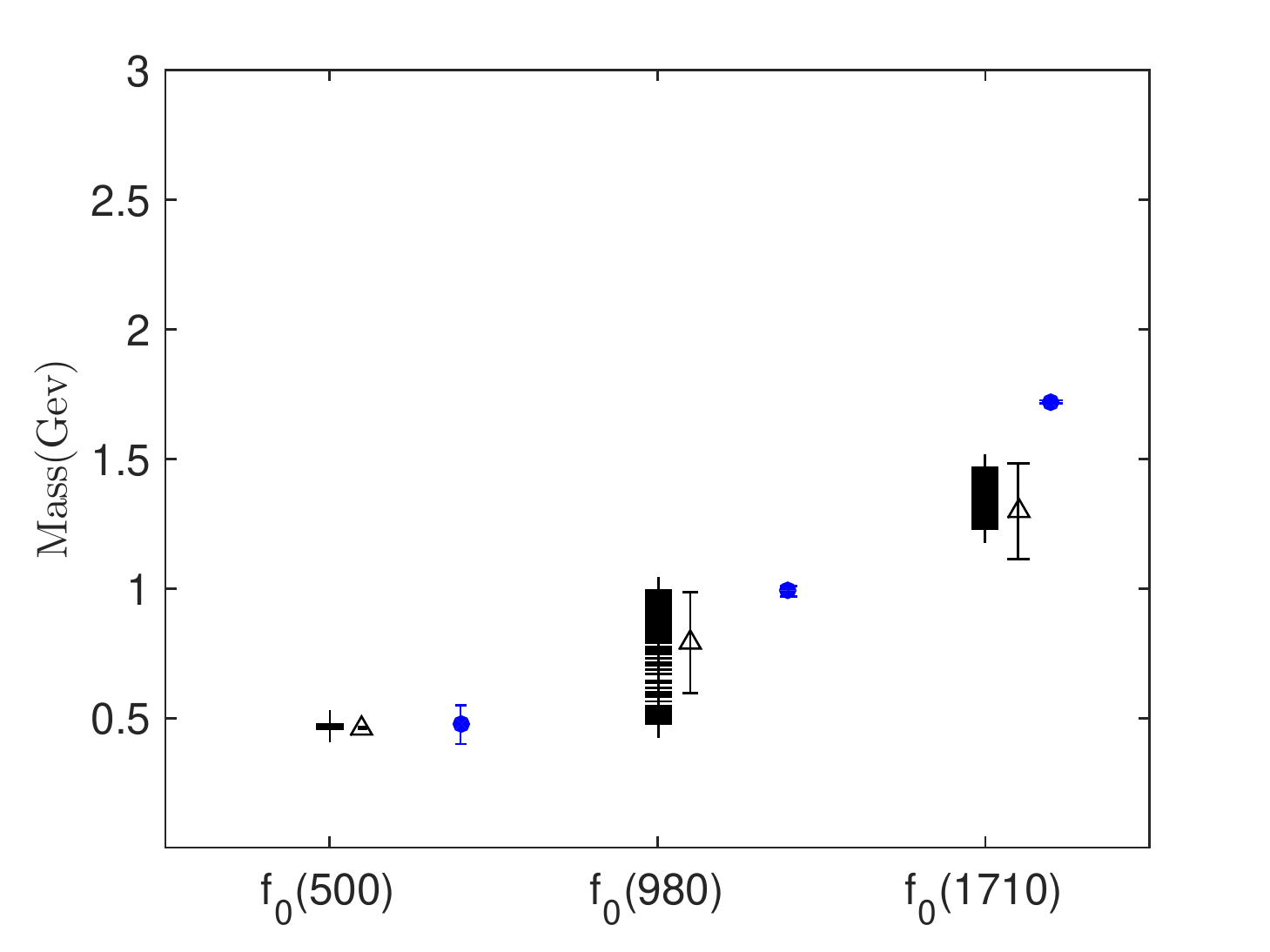}
		\hskip 1cm
		\epsfxsize = 5 cm
		\includegraphics[height=6 cm]{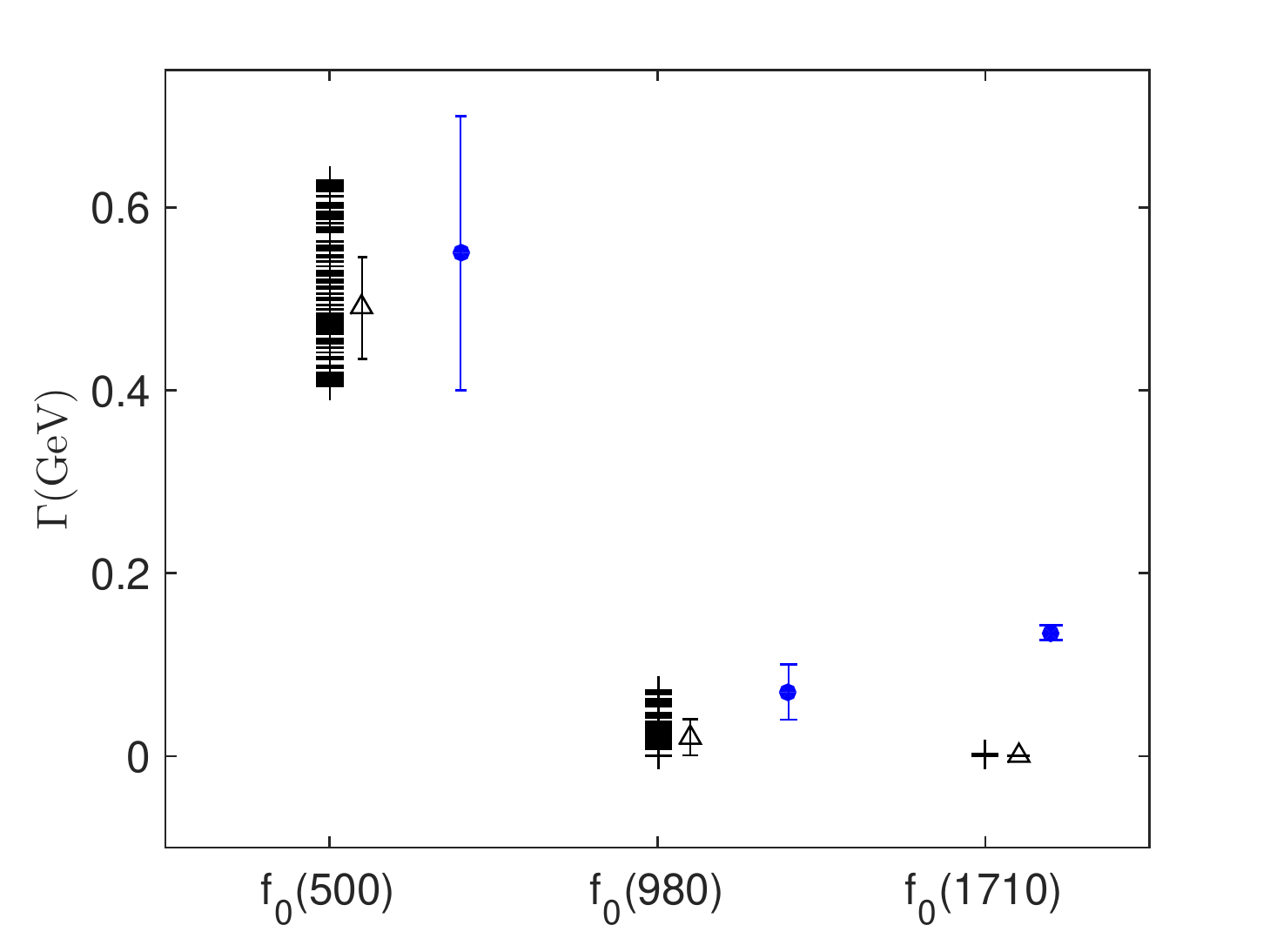}
		\caption{Masses and decay widths of isosinglet scalar mesons obtained from the Monte Carlo simulation for scenario III (pluses) are compared with their experimental values in Table \ref{table1} (solid circles with error bars). In order to compare predictions with the experimental data easier, also the average values (triangles) and  standard deviations around the averages (error bars) are depicted. The masses and decay widths of $f_0(500)$ and $f_0(980)$ are in the experimental range but the predicted decay width of $f_0(1710)$ is too small compared with the experimental data.	\label{fs3}}
		\end{center}
\end{figure}

\begin{figure}[!htbp]
  	\begin{center}
  		\vskip 0.5 cm
  		\epsfxsize = 5 cm
  		\includegraphics[height=6 cm]{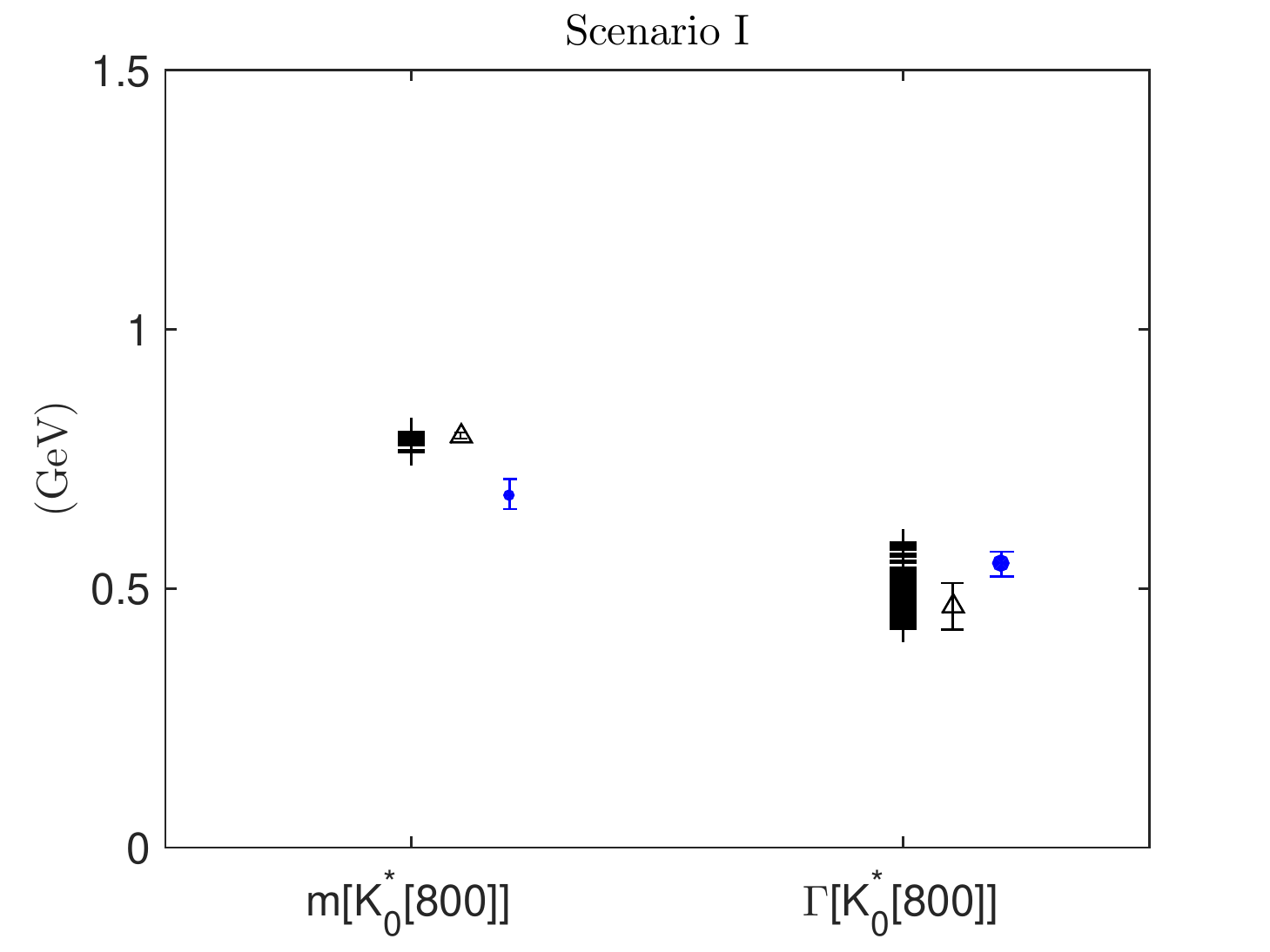}
  		\hskip 1cm
  		\epsfxsize = 5 cm
  		\includegraphics[height=6cm]{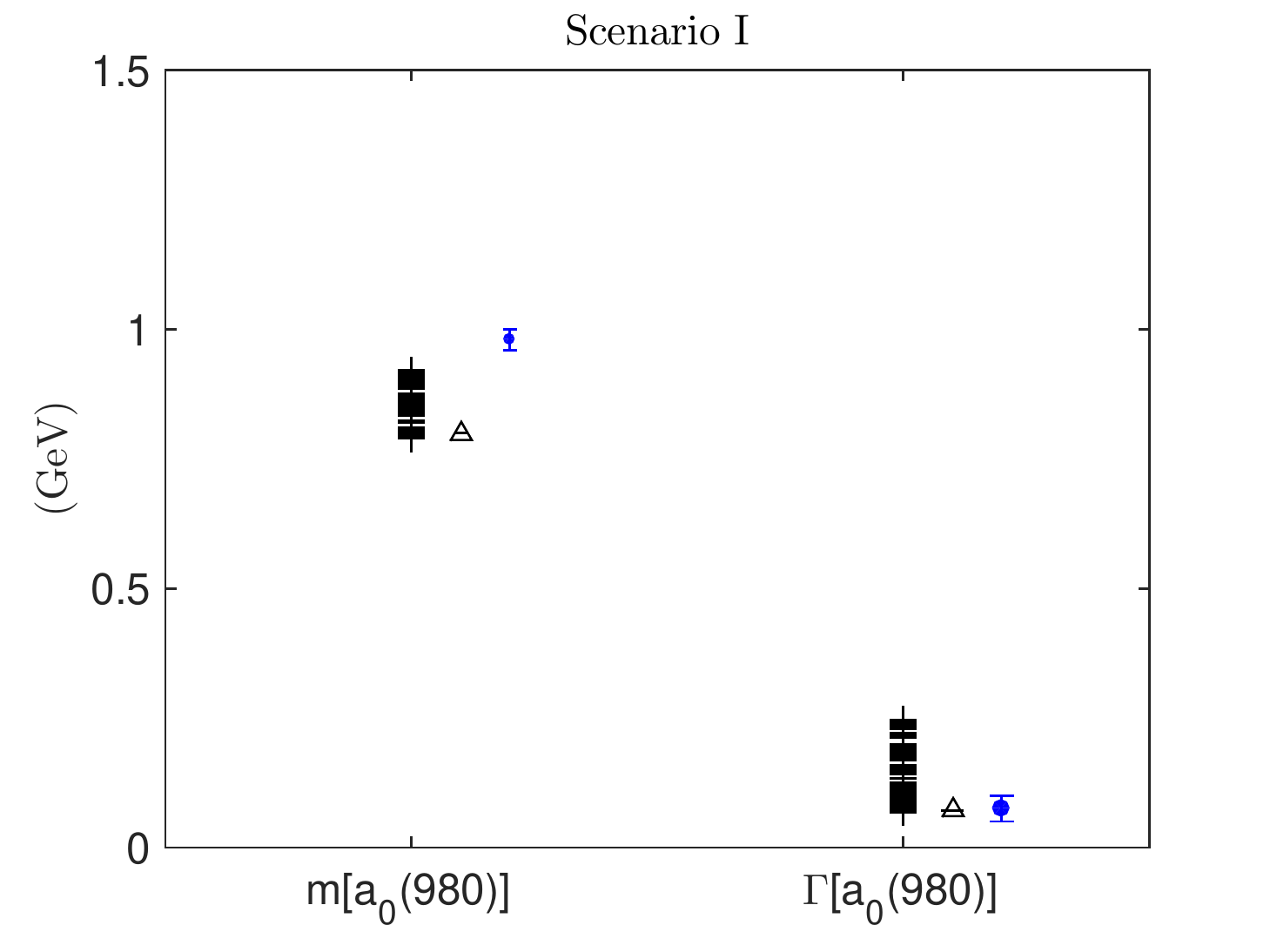}
  		\vskip 0.5 cm
  		\epsfxsize = 5 cm
  		\includegraphics[height=6 cm]{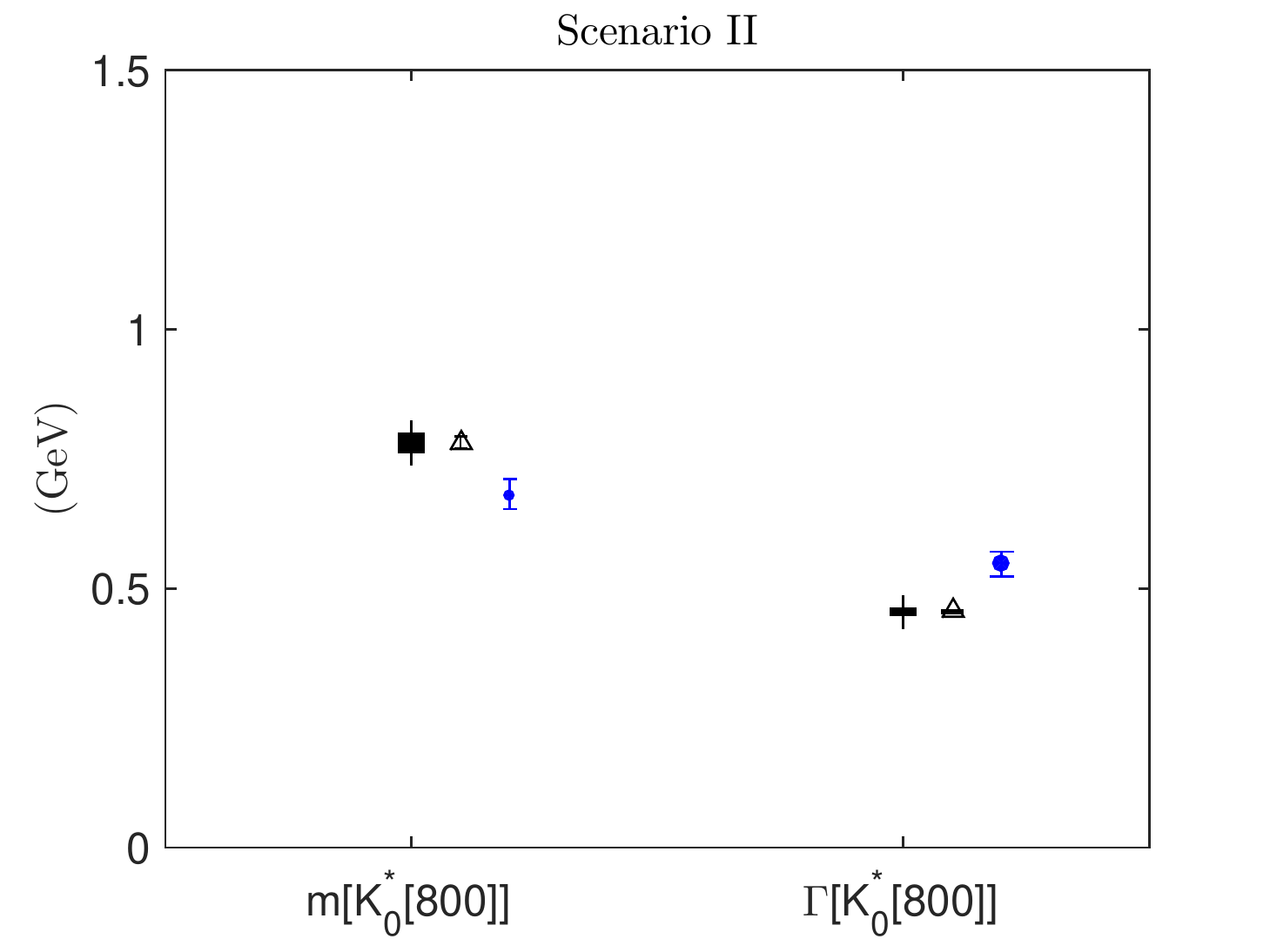}
  		\hskip 1cm
  		\epsfxsize = 5 cm
  		\includegraphics[height=6cm]{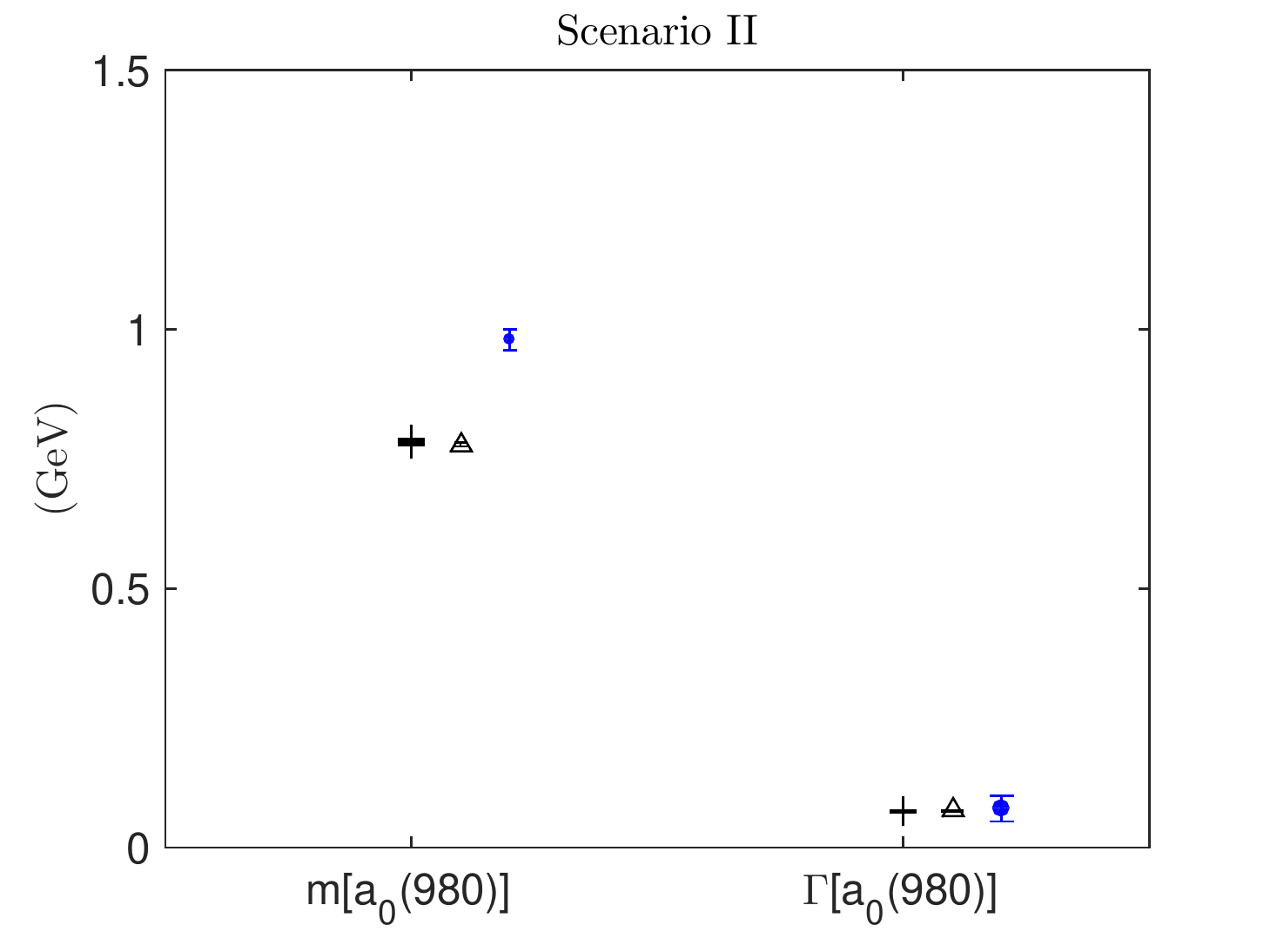}
  		\vskip 0.5 cm
  		\epsfxsize = 5 cm
  		\includegraphics[height=6 cm]{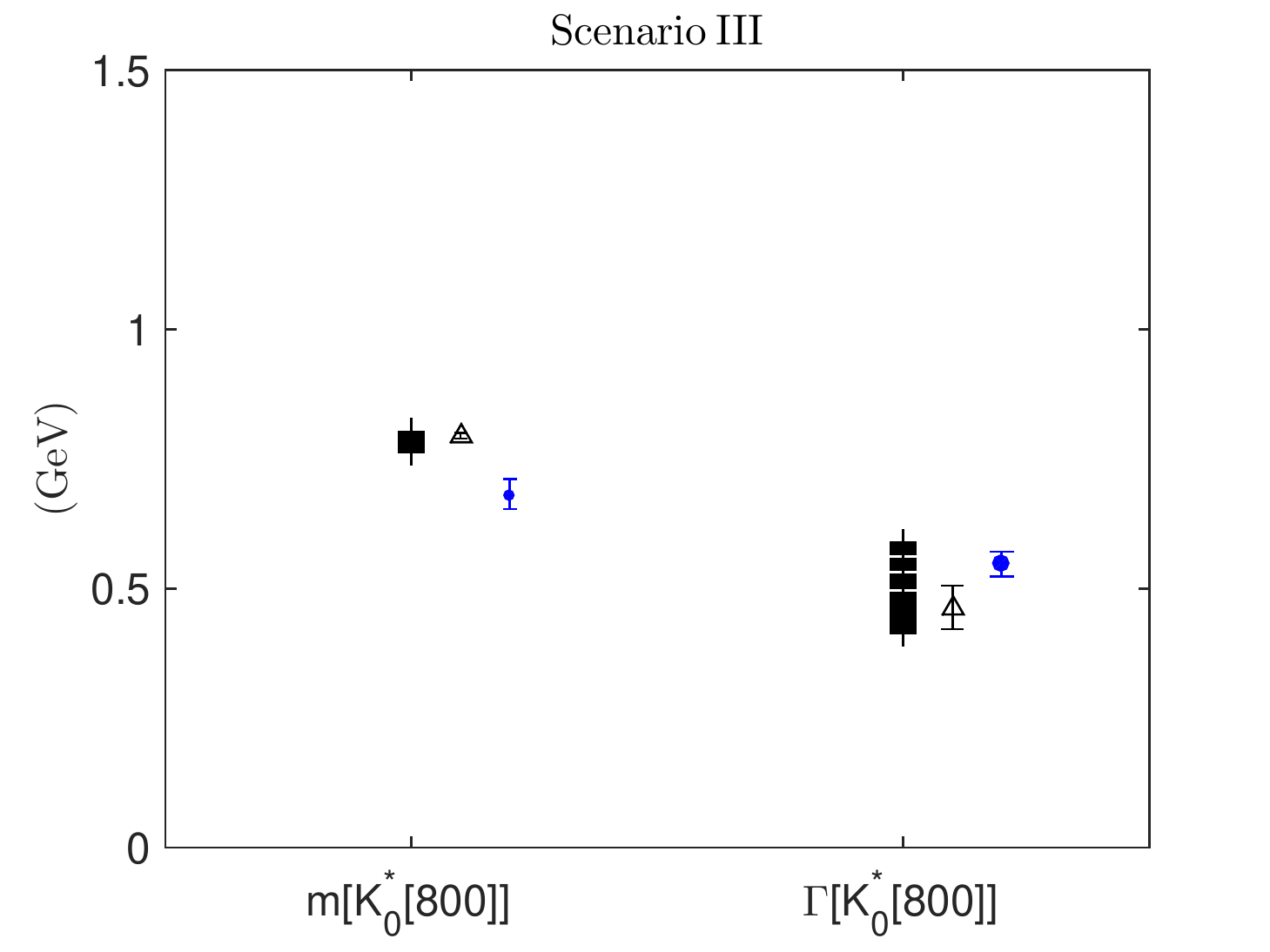}
  		\hskip 1cm
  		\epsfxsize = 5 cm
  		\includegraphics[height=6 cm]{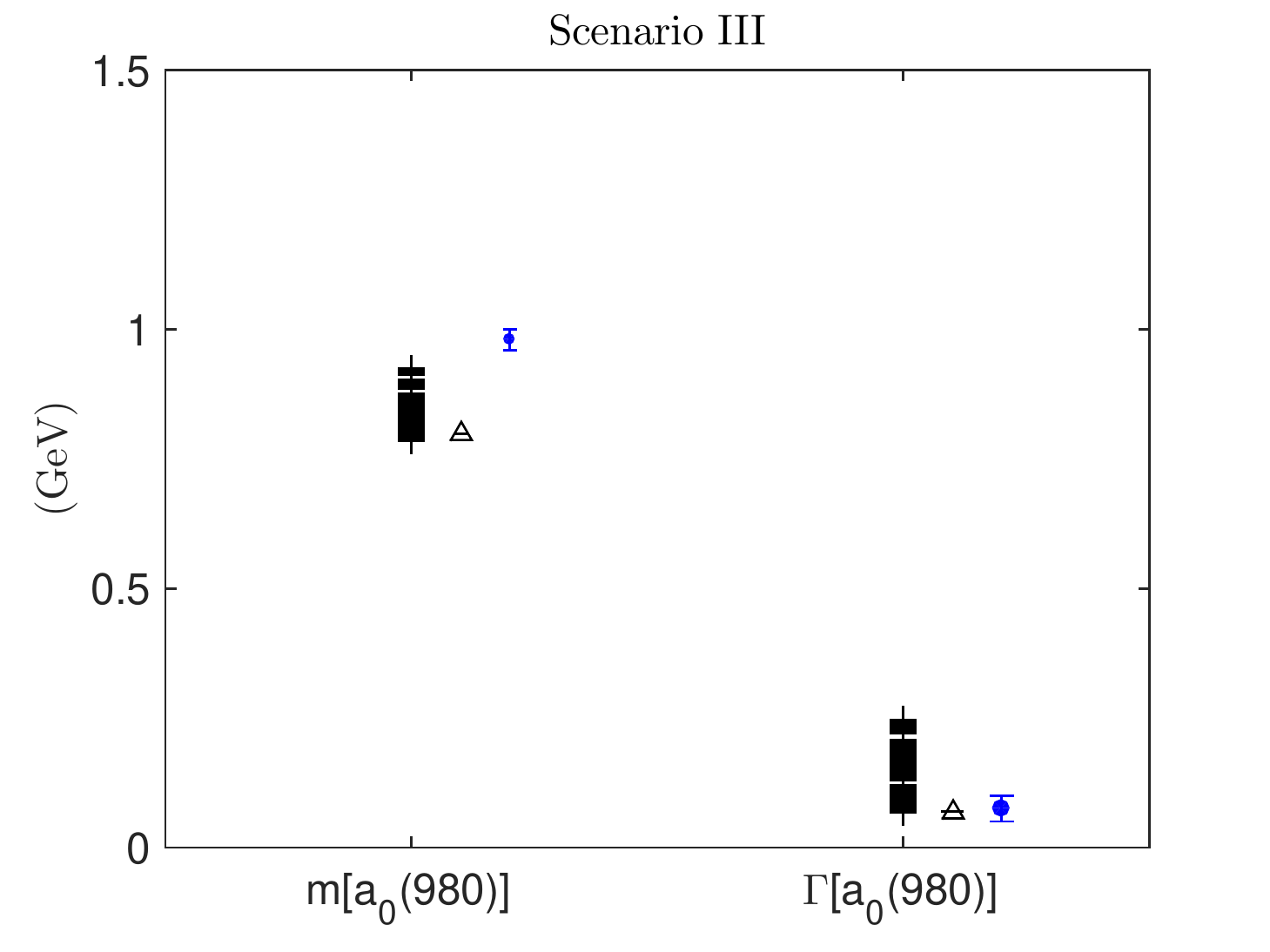}
  		\caption{ Predictions of the model for masses and decay widths of kappa and $a_0(980)$ (pluses) for scenario I (the first row),  II (the middle row) and III (the last row), compared with the experimental inputs presented in Table I (solid circles with error bars). In order to compare predictions with the experimental data easier, also the average values (triangles) and  standard deviations around the averages (error bars) are depicted. The predictions of all three scenarios are very close to each other and similar to SNLSM results. As expected, including scalar glueball, does not have considerable effect on the properties of these mesons. Just for $a_0(980)$, the predicted decay width is in better agreement with experimental range compared with the prediction of SNLSM (Table \ref{table2}).\label{kappa1}}
  	\end{center}
  \end{figure}

\begin{figure}[H]
		\begin{center}
			\epsfxsize = 1 cm
			\includegraphics[scale=0.5]{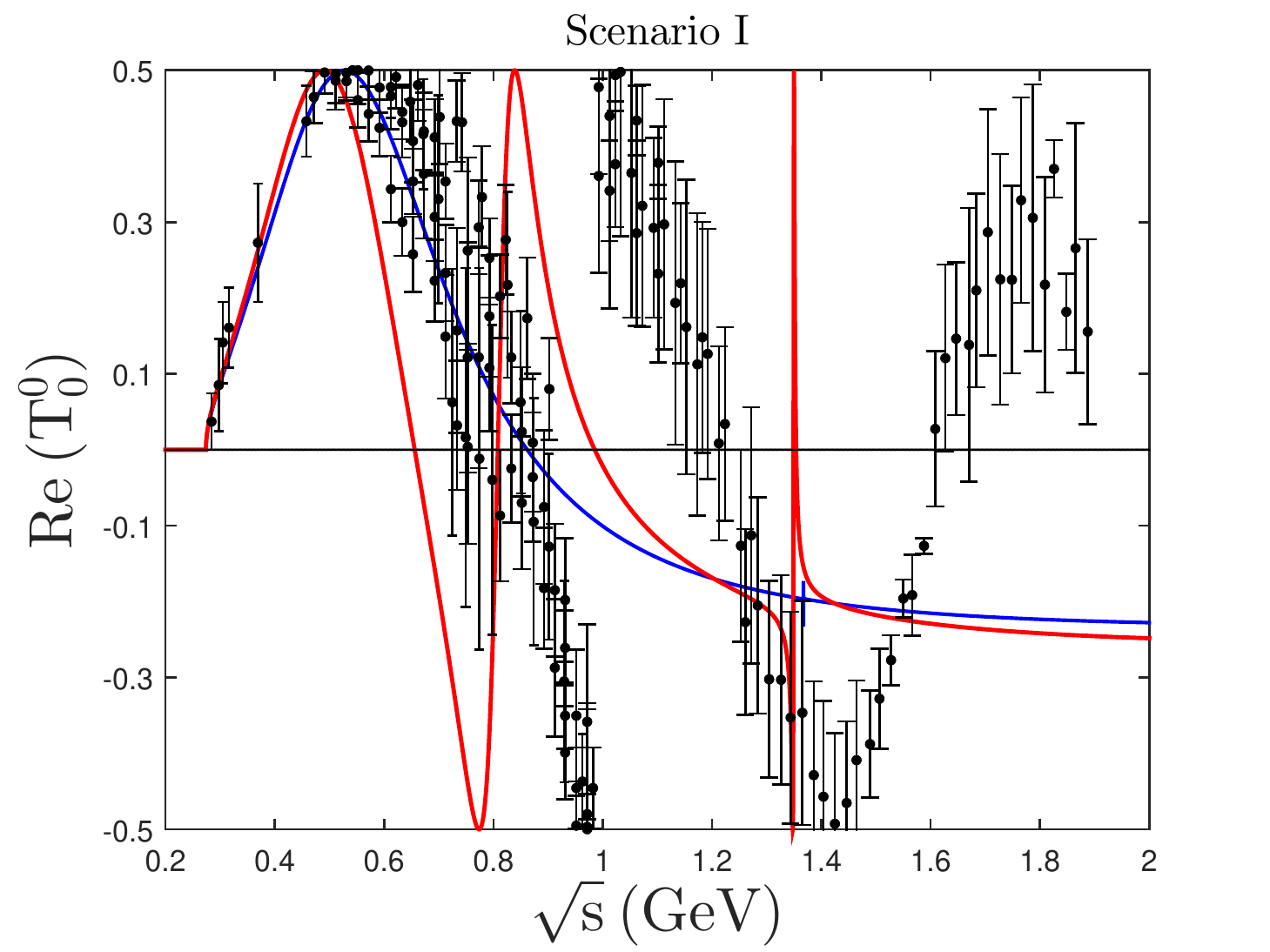}
			\hskip 0.2 cm
			\epsfxsize = 1 cm
			\includegraphics[scale=0.5]{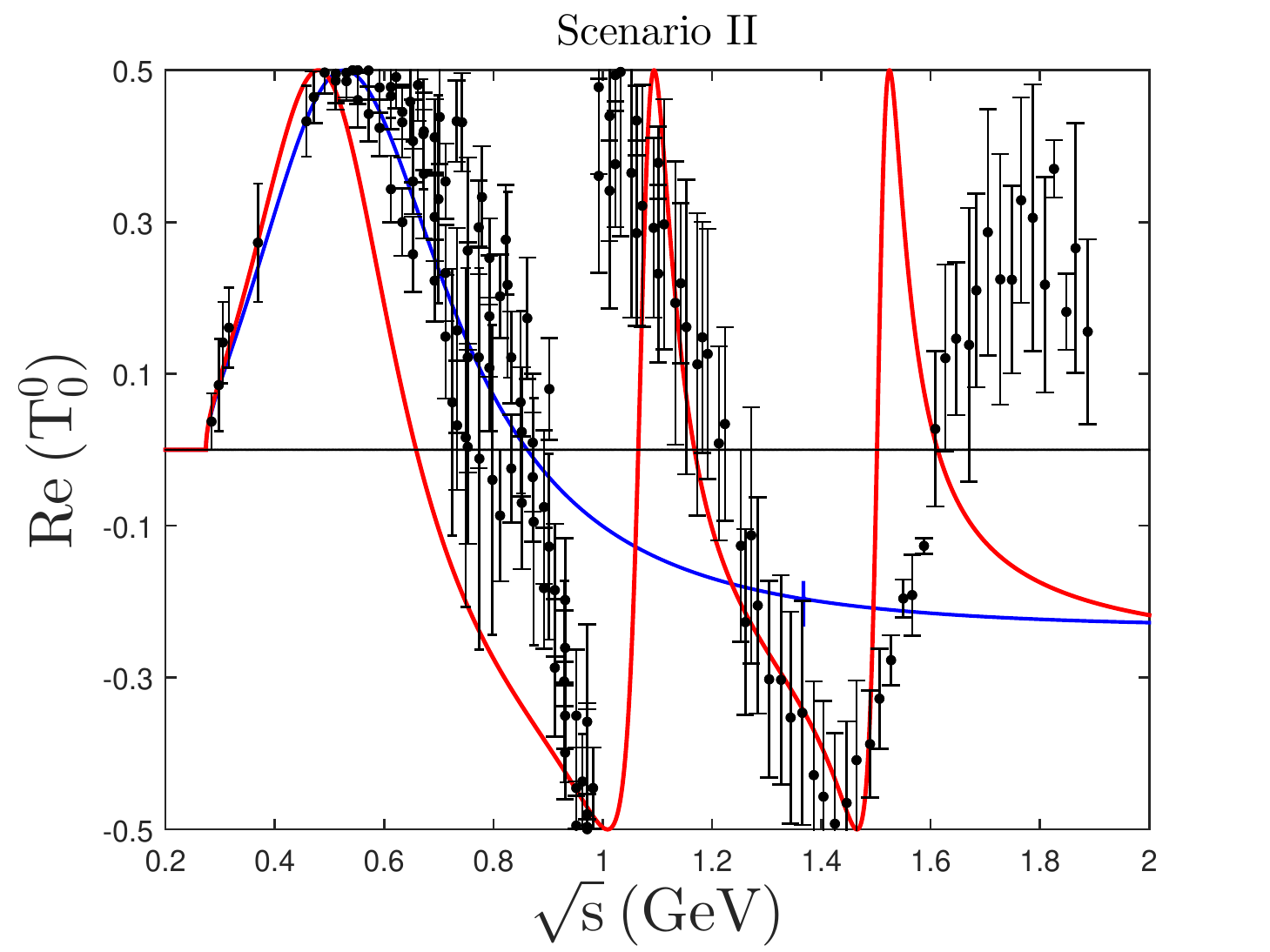}
			\vskip 0.2cm
			\epsfxsize = 1 cm
			\includegraphics[scale=0.5]{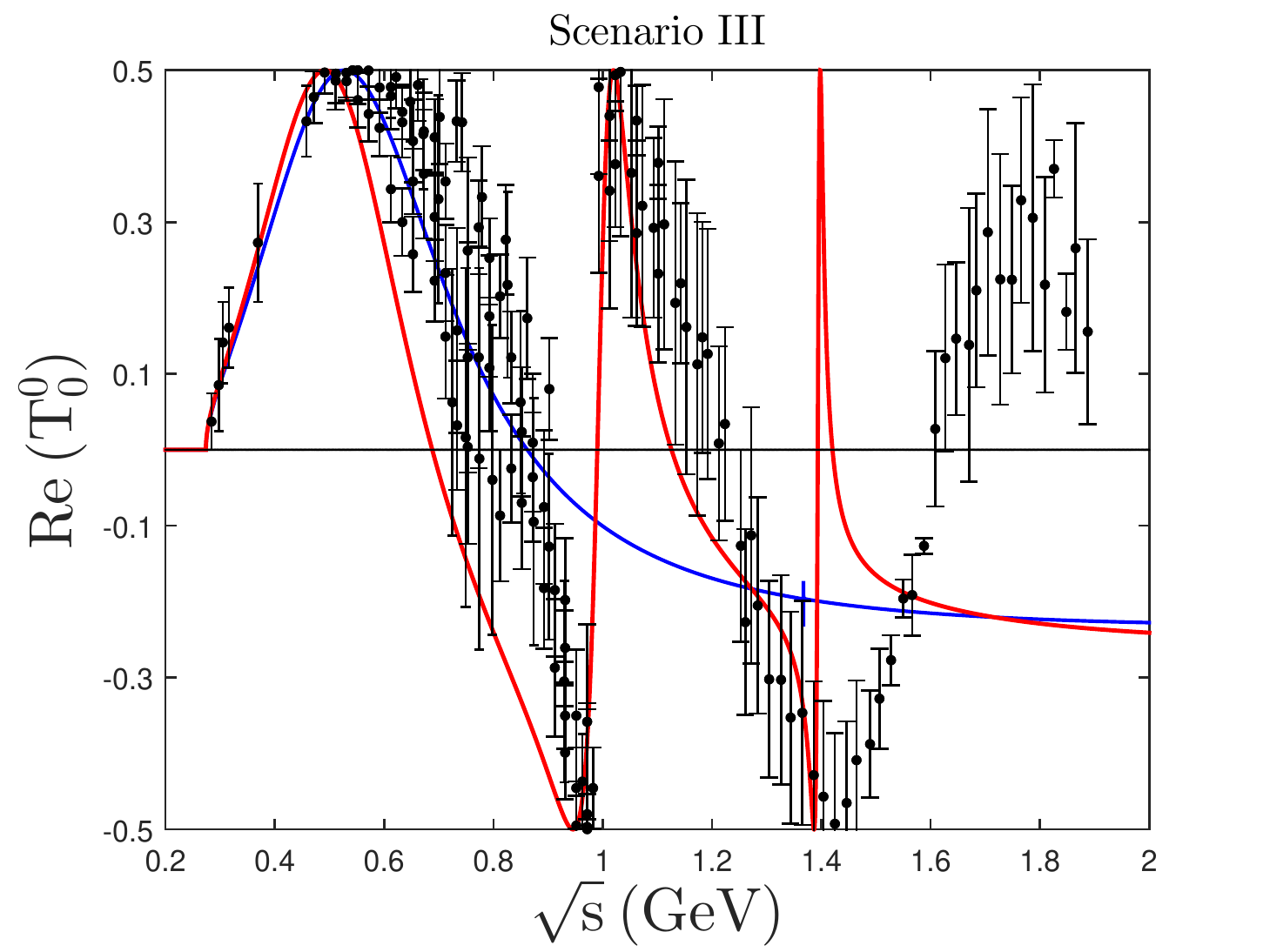}
			\caption{Real part of the K-matrix unitarized $\pi \pi$ scattering amplitude for scenarios I, II, III  (solid line: SNLSM with glueball and dot-dashed line: SNLSM without glueball). While the agreement with experimental data up to about $1$ GeV is almost lost after adding glueball, compared with the case of SNLSM without glueball, the mathematical form of the real part of the amplitude is now in better agreement with experiment for the region above $1$ GeV (especially for scenario II and III).}\label{pi_pi_fig}
		\end{center}
	\end{figure}

\begin{figure}[H]
				\begin{center}
					\epsfxsize = 1 cm
					\includegraphics[scale=0.45]{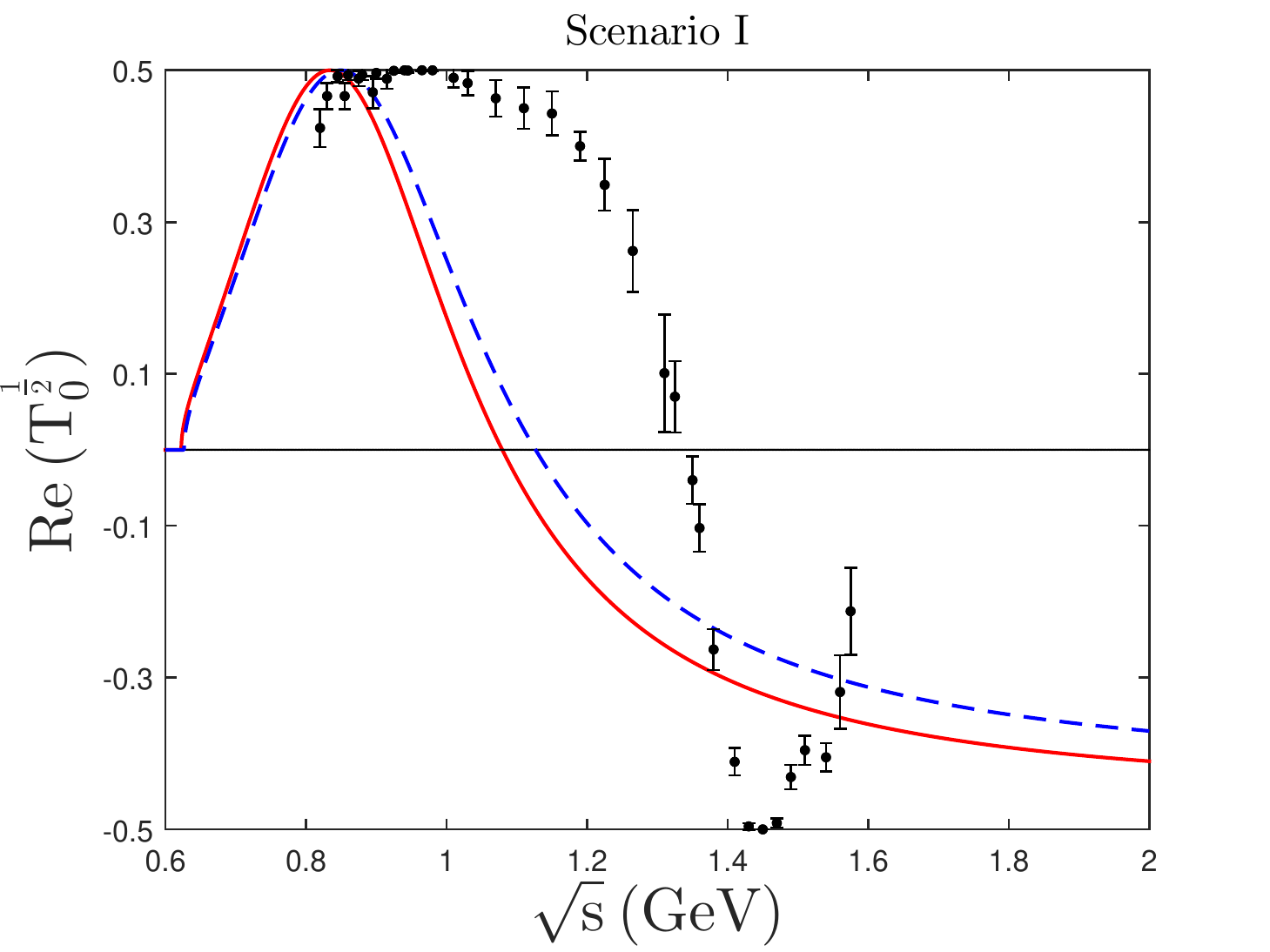}
					\hskip 0.2 cm
					\epsfxsize = 1 cm
					\includegraphics[scale=0.45]{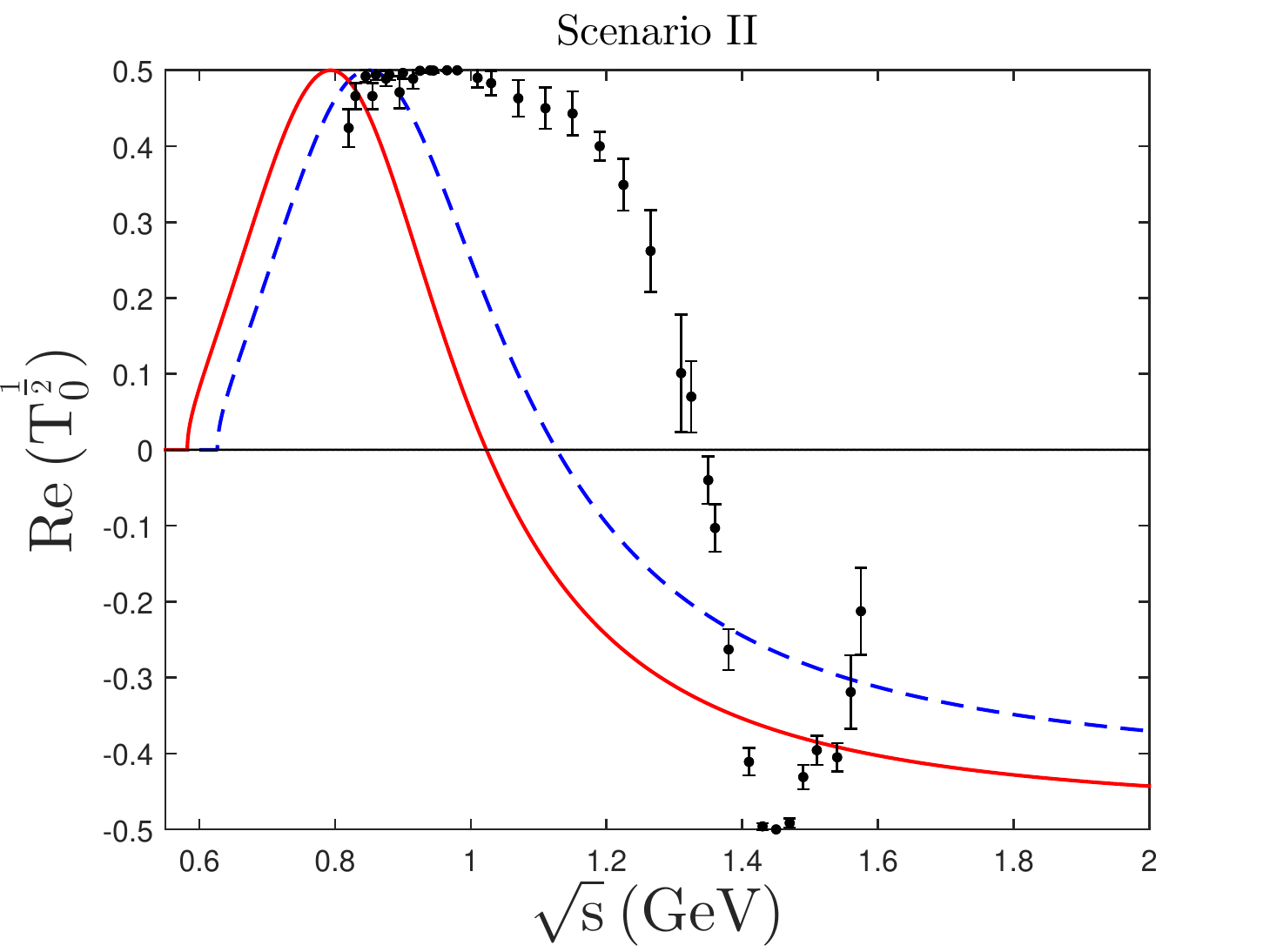}
					\hskip 0.1 cm
					\epsfxsize = 1 cm
					\includegraphics[scale=0.45]{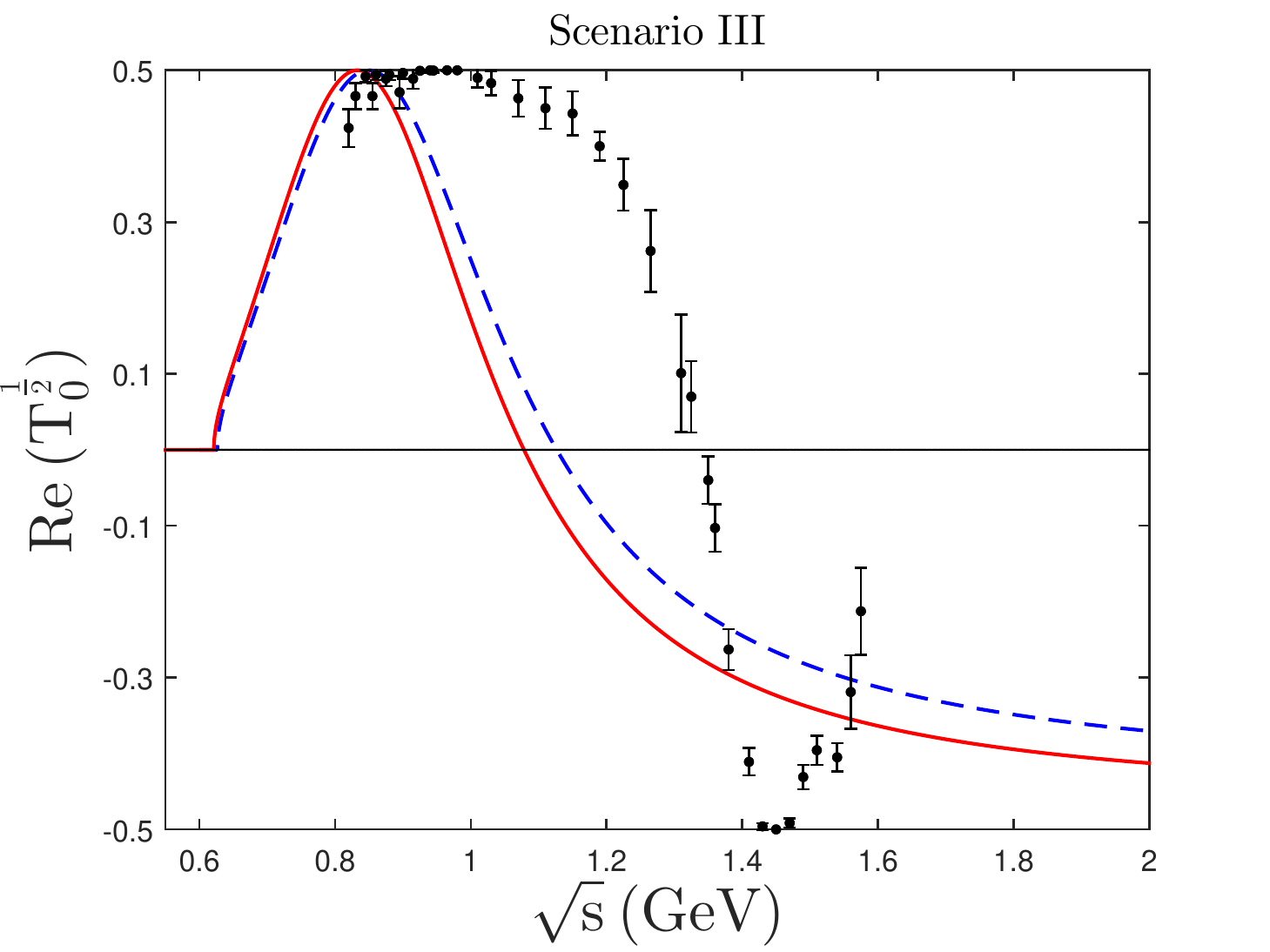}
					\caption{Real part of the K-matrix unitarized $\pi K$ scattering amplitude for scenarios I, II, III  (solid line: SNLSM with glueball and dot-dashed line: SNLSM without glueball). Less agreement with experimental data is achieved in the case of adding glueball (scenario II) compared with the case of SNLSM without glueball. }\label{pi_K_fig}
				\end{center}
			\end{figure}

\begin{figure}[H]
			\begin{center}
				\epsfxsize = 1 cm
				\includegraphics[scale=0.45]{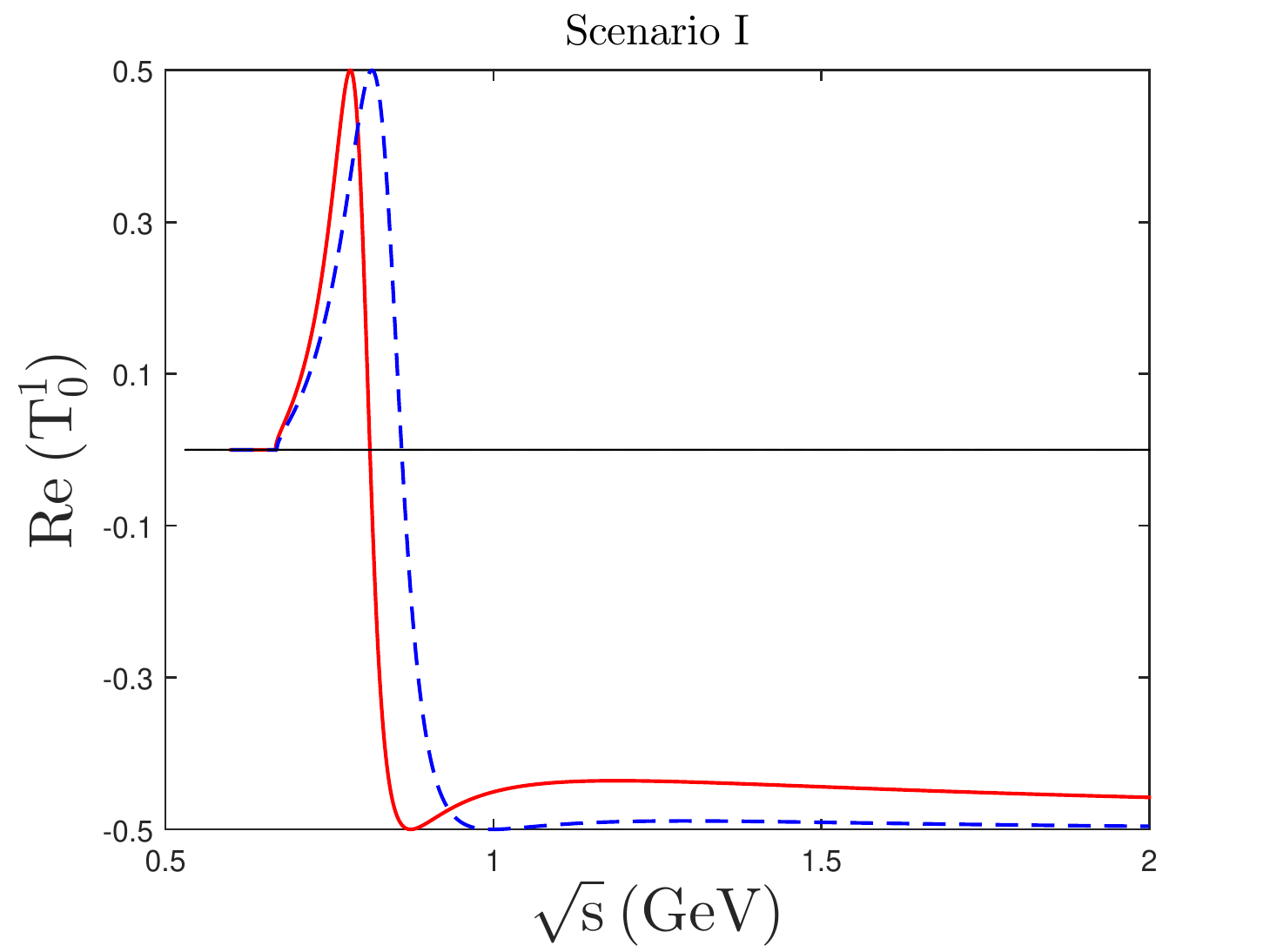}
				\hskip 0.2 cm
				\epsfxsize = 1 cm
				\includegraphics[scale=0.45]{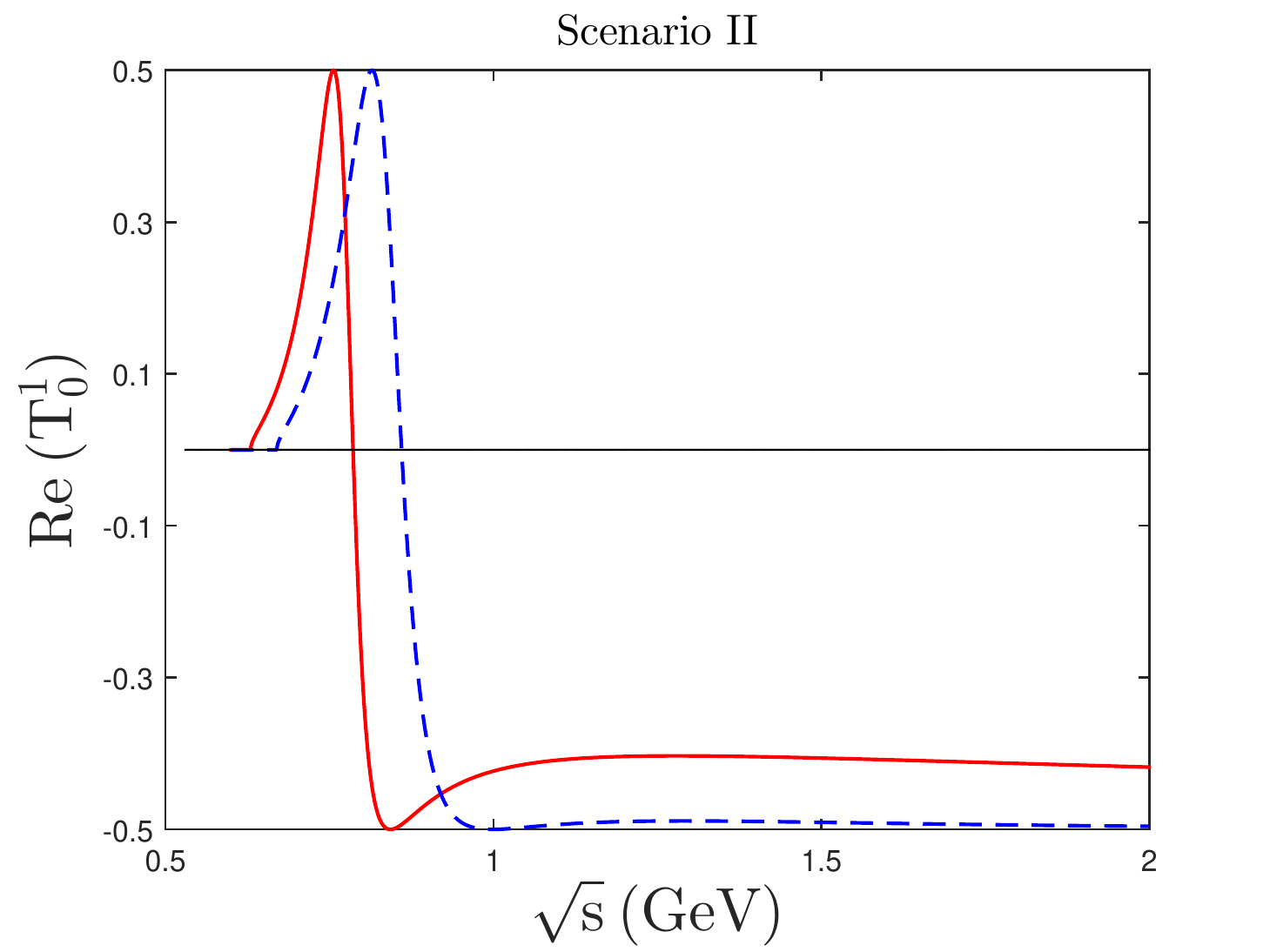}
				\vskip 0.1 cm
				\epsfxsize = 1 cm
				\includegraphics[scale=0.45]{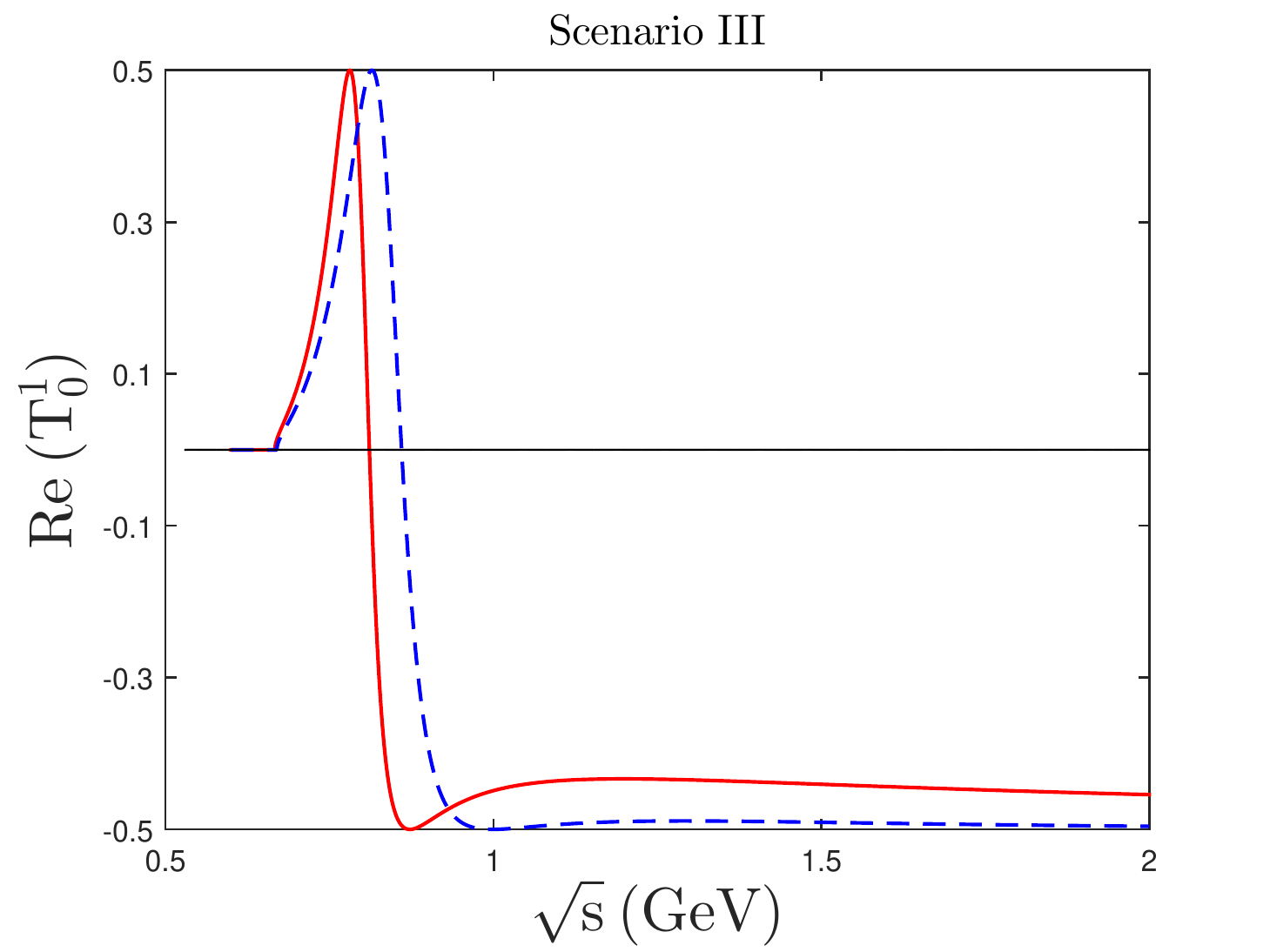}
				\caption{Real part of the K-matrix unitarized $\pi \eta$ scattering amplitude for scenarios I, II, III  (solid line: SNLSM with glueball and dot-dashed line: SNLSM without glueball). As yet there is no experimental data. }\label{pi_eta_fig}
			\end{center}
		\end{figure}

 Therefore, we have a Lagrangian with twelve unknown coefficients $(c_2,c_{4}^{a}, c_{4}^{b}, c_{6}^{a}, c_{6}^b,\gamma, h_{0}, c^{'},\alpha_{1},\alpha_{3}, A_{1}, A_3)$.  
To obtain these unknown parameters, we again apply an iterative Monte-Carlo simulation as in the previous section but now replacing the function $\chi_0$ with the following  $\chi$ function  which should be minimized
 
 \begin{equation}
 \chi=\chi_0+\dfrac{\rvert m_{f_{3}}^{\rm{exp.}}-m_{f_{3}}^{\rm{theo.}}\rvert}{m_{f_{3}}^{exp}}+\dfrac{\rvert \Gamma_{f_{3}}^{\rm{exp.}}-\Gamma_{f_{3}}^{\rm{theo.}}\rvert}{\Gamma_{f_{3}}^{\rm{exp.}}},\nonumber
 \end{equation}
 where  $f_3$ is one of the three isoscalar scalars above 1 GeV, i.e., $f_0(1370)$, $f_0(1500)$ and $f_0(1710)$. This leads to three scenarios which are studied in this work. The predictions of SNLSM in the presence of scalar glueball are presented in Tables \ref{table4}-\ref{table7} and Figs \ref{fs1}- \ref{pi_eta_fig}. Note that for  scenarios I and III, the minimum $\chi$'s obtained from iterative Monte Carlo simulations are greater than their corresponding $(\chi)_{\rm{exp.}}$ which equals $1.954$ for the first and $1.477$ for the third scenario. The reported average values and the standard deviations for these scenarios are obtained from the sets for which $\chi< 3 (\chi)_{\rm{exp.}} $. However, for scenario II, we find sets for which $\chi$'s  are less than  $(\chi)_{\rm{exp.}}(=1.483)$ and the reported averages are calculated over these sets. The typical Lagrangian parameters for this scenario are given in Table \ref{para2} for $\chi=1.1$.

\begin{table}[!htbp]
\footnotesize
\centering
\caption{Typical predicted Lagrangian parameters: $c_2,\,c_4^a,\,c_4^b,\,c_6^a,\,c_6^b,\,c_3,\,c',\,h_0,\,\gamma,\,A_1,\,A_3$ and vacuum parameters: $\alpha_1,\,\alpha_3$ for SNLSM with glueball for $\chi=1.1$ (scenario II).}
\label{para2}
\begin{tabular}{@{}ccccccc@{}}
\noalign{\hrule height 1pt}
\noalign{\hrule height 1pt}
$c_2$& $c_4^a$ & $c_4^b$ &$c_6^a$& $c_6^b$  \\
\noalign{\hrule height 1pt}
$-3.34$ & $17.2$ & $4.27\times 10^{-2}$ & $8.35\times 10^{-2}$ & $3.42\times 10^{-2}$ \\
\noalign{\hrule height 1pt}\\
 $c_3$ &$c'$ & $h_0\,{\rm (GeV})$ & $\gamma$ \\
\noalign{\hrule height 1pt}
  $-1.08\times 10^{-1}$&$-2.36\times 10^{-2}$&$7.73\times 10^{-2}$&$3.78\times 10^{-1}$\\
\noalign{\hrule height 1pt}\\
$A_1 \,{\rm (GeV}^3)$ &$A_3 \,{\rm (GeV}^3)$ & $\alpha_1 \,{\rm (GeV})$& $\alpha_3 \,{\rm (GeV})$\\
\noalign{\hrule height 1pt}
$6.15\times 10^{-4}$ & $1.51\times 10^{-2}$& $6.55\times 10^{-2}$& $9.32\times 10^{-2}$\\				 
\noalign{\hrule height 1pt}
\end{tabular}
\end{table}
		
In Tables \ref{table4}-\ref{table6}, the percentages of quark and glue components  of the lowest lying isoscalars besides the ones for $f_0(1370)$, $f_0(1500)$ and $f_0(1710)$ are displayed for the three scenarios. Comparing these three tables, we can see that for scenario II (Table \ref{table5}), $77.36$ \%  of $f_0(1500)$ is made of glue component which shows $f_0(1500)$ can be considered as the  scalar glueball. Also for this scenario, the strange component of  $f_0(980)$
is $80.10$\% and the light quark percentage of $f_0(500)$ is about $85.74\%$. As a matter of fact, due to the quark model and also the strong coupling of $f_0(980)$ to kaons, the structure of $f_0(980)$ can be considered as a pure strange quarkonium $ s\bar{s}$ \cite{8reff0,6reff0, 5reff0} and $f_0(500)$ as a $(u\bar{u}+d\bar{d})/\sqrt{2}$ state\cite{uu+dd_sigma}. The predictions of the first and the third scenarios which show a large glueball component for $\sigma$ and $f_0(980)$ are not consistent with the common interpretation of them as quark states ($f_0(980)$ is a multiquark \cite{21soode2,qqqq} or $K \bar{K}$ bound state \cite{kkmolecule} with sizable $s$-quark content and $\sigma$ is a multiquark with considerable $u-d$ component \footnote{The multiquark or molecule structures for isosinglet scalars can not be predicted in our model, which only contains $q\bar{q}$ nonet. Therefore, our model predicts $f_0(980)$ and $f_0(500)$ as predominately $s\bar{s}$ and  $(u\bar{u}+d\bar{d})/\sqrt{2}$,  respectively.}).

\begin{table}[!htbp]
	\footnotesize
	\centering
	\caption{The estimate of the model for the percentages of quark and glue components of isoscalars for scenario I.}
	{\begin{tabular}{@{}l|llll@{}}
			\noalign{\hrule height 1pt}
			\noalign{\hrule height 1pt}
			& $\frac{u\bar{u}+d \bar{d}}{\sqrt{2}}$ & $s\bar{s}$& $G$  \\
			\noalign{\hrule height 1pt}
			$f_0(500) $ &   $ 45.80 \pm 35.38$ & $0.56 \pm 0.51$& $53.63 \pm 35.29$ \\
			$f_0(980)$ &   $ 54.08 \pm  35.44$ & $ 2.36\pm 2.32$ & $43.54 \pm 33.65$ \\
			$f_0(1370) $  &  $0 0.11\pm 0.12$ & $97.06 \pm 2.50$ & $2.82 \pm2.39$   \\
			\noalign{\hrule height 1pt}
			\noalign{\hrule height 1pt}
		\end{tabular}\label{table4}}
\end{table}

\begin{table}[!htbp]
	\footnotesize
	\centering
	\caption{The estimate of the model for the percentages of quark and glue components of isoscalars for scenario II.}
	{\begin{tabular}{@{}l|llll@{}}
			\noalign{\hrule height 1pt}
			\noalign{\hrule height 1pt}
			& $\frac{u\bar{u}+d \bar{d}}{\sqrt{2}}$ & $s\bar{s}$& $G$  \\
			\noalign{\hrule height 1pt}
			$f_0(500) $ &   $85.74 \pm 0.59$ & $3.05 \pm 0.23$& $11.20 \pm 0.80$ \\
			$f_0(980)$ &   $ 8.46 \pm 0.48$ & $80.10 \pm 2.72$ & $11.42 \pm 2.25$ \\
			$f_0(1500) $  &  $5.79 \pm 0.13$ & $ 16.84 \pm2.94$ & $77.36 \pm  3.03$   \\
			\noalign{\hrule height 1pt}
			\noalign{\hrule height 1pt}
		\end{tabular}\label{table5}}
\end{table}

\begin{table}[!htbp]
	\footnotesize
	\centering
	\caption{The estimate of the model for the percentages of quark and glue components of isoscalars for scenario III.}
	{\begin{tabular}{@{}l|llll@{}}
			\noalign{\hrule height 1pt}
			\noalign{\hrule height 1pt}
			& $\frac{u\bar{u}+d \bar{d}}{\sqrt{2}}$ & $s\bar{s}$& $G$  \\
			\noalign{\hrule height 1pt}
			$f_0(500) $ &   $51.52 \pm 34.74$ & $0.66 \pm  0.59$& $47.81 \pm 34.78$ \\
			$f_0(980)$ &   $48.13 \pm 34.96$ & $5.17 \pm 7.51$ & $46.68 \pm 32.03$ \\
			$f_0(1710) $  &  $0.34 \pm 0.62$ & $94.15\pm 7.94$ & $ 5.50 \pm  7.33$   \\
			\noalign{\hrule height 1pt}
			\noalign{\hrule height 1pt}
		\end{tabular}\label{table6}}
\end{table}

In Table \ref{table7},  the predictions of the model for the masses  and decay constants of pseudoscalar mesons are given. As we expected, the results are almost the same as those presented in Table \ref{table3}; the scalar glueball should not affect the pseudoscalar sector of the Lagrangian. 
\begin{table}[!htbp]
	\footnotesize
	\centering
	\caption{{Prediction of SNLSM in the presence of the scalar glueball for the masses and decay constants of pseudoscalars below 1 GeV for scenario II.}}
	{\begin{tabular}{@{}ll|ll@{}}
			\noalign{\hrule height 1pt}
			\noalign{\hrule height 1pt}
			& Mass (GeV) &   Decay constant (GeV) \\
			\noalign{\hrule height 1pt}
			$\pi $ &   $0.137 \pm 2.67 \times 10^{-5} $ &  $0.131 \pm 5.5\times 10^{-5}$ \\
			$K$ &   $0.448 \pm 0.005$ &  $0.158 \pm 3.86\times 10^{-4} $ \\
			$\eta $  &  $ 0.496 \pm  0.005$ & \quad  \\
			$\eta \prime $  &  $0.985 \pm 0.002$ & $\quad$  \\
			\noalign{\hrule height 1pt}
			\noalign{\hrule height 1pt}
		\end{tabular}\label{table7}}
\end{table}

As it is seen from Figs. \ref{fs1}-\ref{fs3}, the predictions of the model for the masses and decay widths of $f_0(500)$ and $f_0(980)$ are in experimental range or very close to it.  However, only for scenario II the decay width of the third scalar, i.e., $f_0(1500)$ does match with the experimental range. 
This again confirms that $f_ 0(1500)$ is a preferred candidate 
for the scalar glueball.

 It is clear from Fig. \ref{kappa1} that for all scenarios the average predicted masses and decay widths for $K_0^*(800)$ and also the average predicted masses for $a_0(980)$  are not in the experimental ranges but close to them. As it was expected, the masses and decay widths of these mesons have not been considerably affected by adding the scalar glueball in the Lagrangian and therefore for this case none of the scenarios is simply preferable to another. Just for $a_0(980)$, the predicted decay width is in better agreement with experimental range compared with the prediction of SNLSM (Table \ref{table2}).

Also, we have plotted the predictions of the model for the K-matrix unitarized $\pi\pi$ scattering amplitude for three scenarios for typical values of $\chi$ to see if a better agreement with experiment is obtained. From Fig. \ref{pi_pi_fig} it is clear that while the agreement with experimental data up to $1$ GeV is almost lost compared with the case of SNLSM without glueball (Fig. \ref{plotampwg}), the mathematical form of the real part of the amplitude is now analogous to experiment up to about $2$ GeV (it is more clear for scenarios II and III). For scenarios II and III, not only the similarity in mathematical structure is seen, but also there are some regions above $1$ GeV for which the curve of predictions goes through the experimental range. Motivated by this similarity in shape, we are encouraged to follow the same procedure and implement glueball in GLSM which does not show good agreement with experimental data above $1$ GeV. 
\par 
 Finally, it is shown in Fig. \ref{pi_K_fig} that adding the scalar glueball shifts the K-matrix $\pi K$ scattering amplitude slightly for scenarios I and III and considerably for scenario II in the sense that in this case less agreement with experimental data is achieved for scenario II compared with the case of SNLSM without glueball. Note that adding glueball does not affect the mathematical form of the K-matrix unitarized scattering amplitudes of $\pi K$  nor the $\pi \eta$ (Fig. \ref{pi_eta_fig}) in contrast to $\pi\pi$.

	\noindent
	
\section{Summery and Conclusion}\label{sec4}
In this paper, we have studied the effect of adding a scalar glueball to the SNLSM to study the properties of scalar and pseudoscalar mesons. The glueball has been mixed with the matter field to improve the results  of SNLSM. It has been shown that among the three different scenarios for choosing the scalar glueball, i.e., $f_0(1370)$, $f_0(1500)$ and $f_0(1710)$,  $f_0(1500)$ is an appropriate candidate (scenario II). The predictions of SNLSM with and without scalar glueball are presented in Table \ref{table_conclusion}. It can be seen that the predictions with glueball (second column)  have been improved compared with the ones without glueball (first column). Since $\gamma_{f_2 \pi \pi}$   is very close to zero  in SNLSM without glueball, therefore the decay width of $f_0(980)$ can not be determined in this model, but after adding the scalar glueball, not only we get $0.07$ GeV for its decay width which agrees with experimental data but also its mass reduces from $1.362$ GeV to $1.066$ GeV which indeed is in better agreement with experiment.  Moreover, The  decay width of $a_0(980)$  gets the value $0.070$ GeV which is a better prediction. 
\par
\begin{figure}[!htbp]
	\begin{center}
          	\epsfxsize = 1 cm
			\includegraphics[scale=0.57]{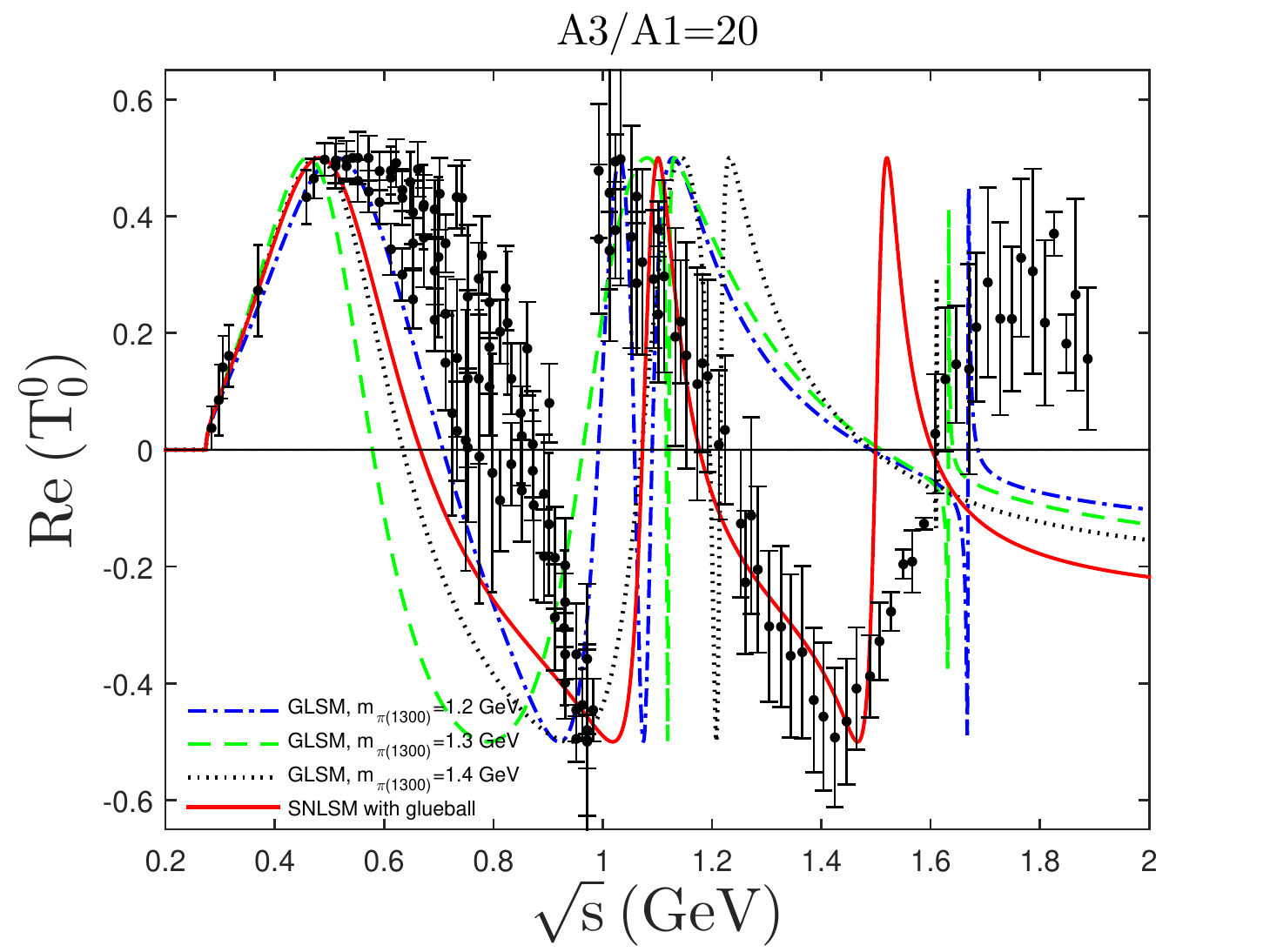}
			\hskip 0.2 cm
			\epsfxsize = 1 cm
			\includegraphics[scale=0.57]{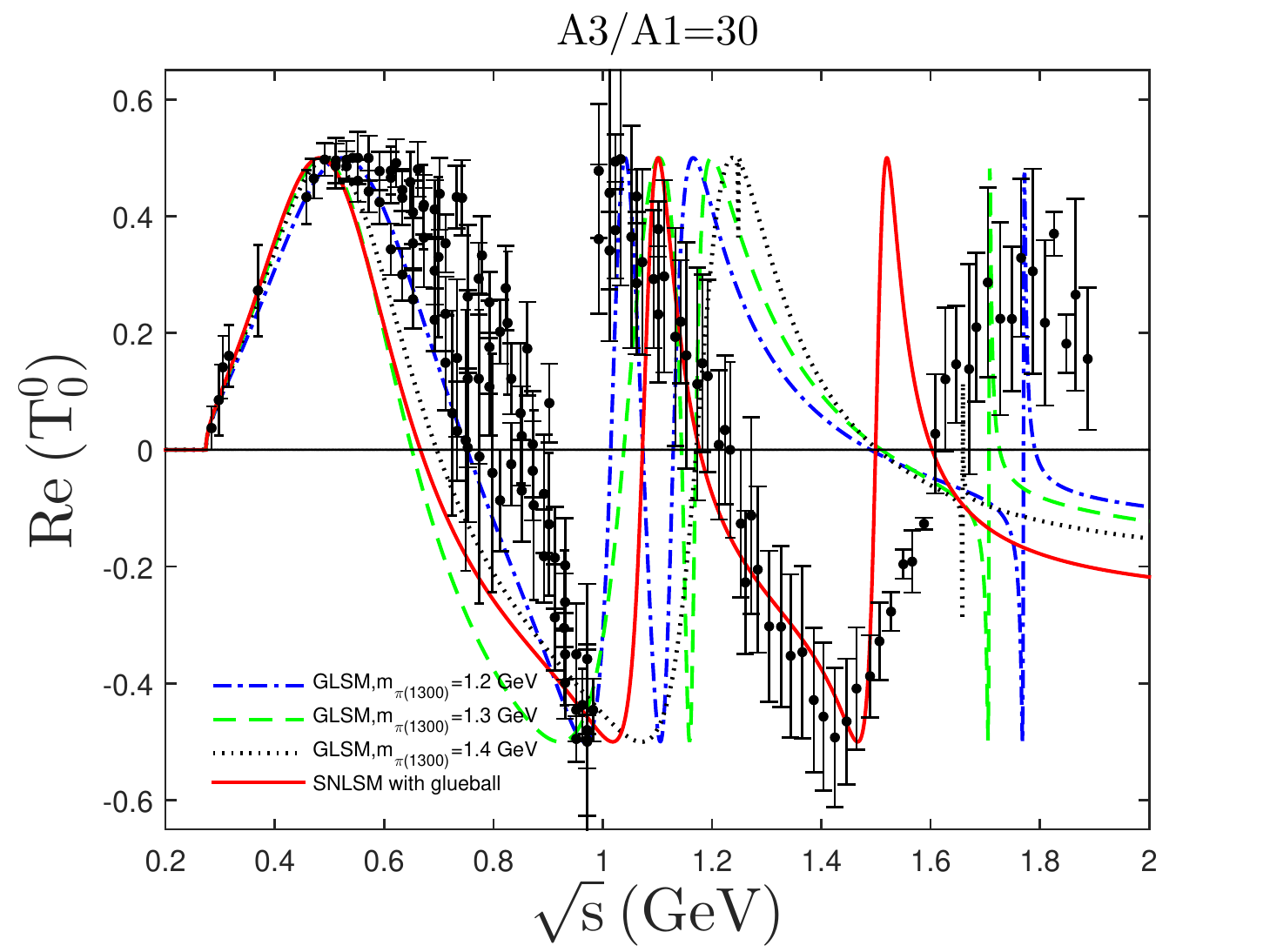}
		\caption{A comparison of the real part of the K-matrix unitarized $\pi \pi$ scattering amplitude of our model (scenario II - for the set with $\chi=1.19$) and the generalized linear sigma model (GLSM) for three different choices of $m[\pi(1300)]$ and $A_3/A_1=20,\,30$ \cite{40soode2,soode2}. For above $1$ GeV, the inclusion of glueball in SNLSM obviously improves  the matching between the prediction of our model and the experimental data. }\label{pipiscat}
		\end{center}
	\end{figure}

\begin{figure}[!htbp]
		\begin{center}
	          	\epsfxsize = 1 cm
				\includegraphics[scale=0.8]{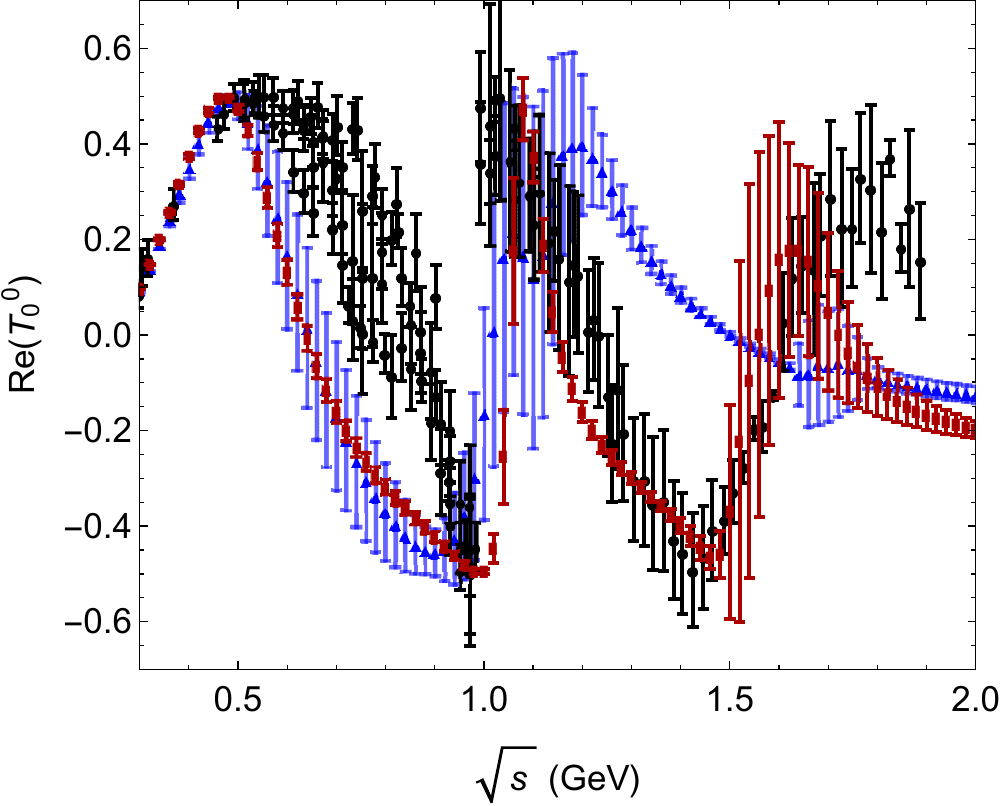}
			\caption{The averages (blue triangles) and standard deviations (error bars) of  the prediction of GLSM for the real part of the K-matrix unitarized $\pi\pi$ scattering amplitude is compared with the average values (red squares) together with standard deviations (error bars) of SNLSM with glueball. Up to about $1.1$ GeV, the results coincide, although in some regions matching with experiment (black dots with error bars) is not seen.
			 It is also clear that while the predictions of SNLSM with glueball for the region of $1.1-1.7$ GeV, do match with experimental data, GLSM misses agreement with data. }\label{pipiscatmean}
			\end{center}
		\end{figure}

\begin{figure}[!htbp]
		\begin{center}
	          	\epsfxsize = 1 cm
				\includegraphics[scale=0.7]{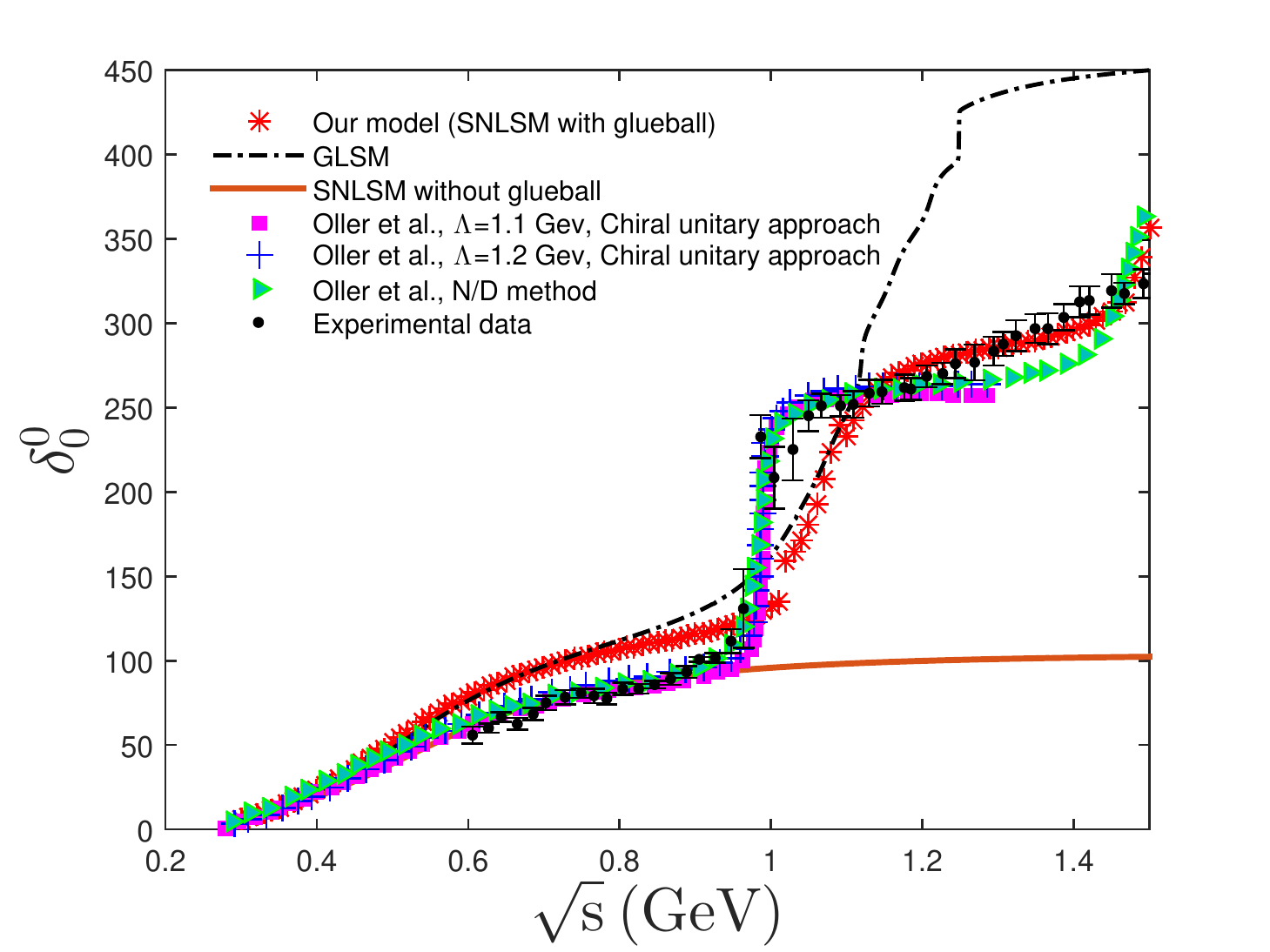}
			\caption{The predictions of SNLSM for elastic $\pi\pi$ scattering phase shift with and without glueball resulted from unitarizing with the K-matrix method are compared with the predictions of GLSM with the same unitarization approach (averaged on different values of $A_3/A_1$ and $m[\pi(1300)])$ \cite{40soode2}. Also the predictions of LO ChPT  resulted from two  unitarizing approaches are depicted: chiral unitary approach (for two choices of cutoff energy $\Lambda=1.1$ and $1.2$ GeV)\cite{ollercua,pelaezrev} and the N/D method \cite{ollernd,pelaezrev}. While up to about $1$ GeV, the prediction of SNLSM without glueball match well with experiment, the prediction of SNLSM including glueball (scenario II) follows the behavior of data up to about $1.5$ GeV and the agreement with data is rather good for $\sqrt{s}=1.1-1.5$ GeV. It is clear that for the energy region below $1.1$ GeV, the predictions of GLSM and our model are close and far from experiment but for $\sqrt{s}>1.1$ GeV, our model is consistent with observed data. Also it can be seen that the predictions of LO ChPT with different unitarization approaches are successful up to values of $\sqrt{s}=1.2$ GeV and from there on a fair description of data is achieved. Note that at this stage it does not make sense to compare the predictions of our model and \cite{ollernd,ollercua} since the coupled channels are absent in our calculations.}\label{pipiscatph}
			\end{center}
		\end{figure}

 
Furthermore, it would be very interesting to compare our results to those obtained  in generalized linear sigma model (GLSM) \cite{f_global,40soode2,soode2}, where mixing among $q\overline{q}$ and $qq\overline{q}\overline{q}$ has been performed with two nonets of scalars and two nonets of pseudoscalars below and above
$1$ GeV. We would expect the predictions of GLSM to be in better agreement with experiment compared with the predictions of SNLSM without glueball. This is made evident by comparing the first column of  Table \ref{table_conclusion} with the first column of Table \ref{two_body2}. It is interesting that adding a scalar glueball to SNLSM does greatly modify this model to predict masses and decay widths in the experimental range or  close to it (See Table \ref{table_conclusion} ). Although  our model lacks the second meson nonet above $1$ GeV, its predictions for masses and widths are comparable with the results of GLSM which is believed to have a better structure of predicting the low energy QCD. The comparison between these two models is given in Table  \ref{two_body2} and Fig \ref{pipiscat}.  It is observed from Table \ref{two_body2} that the decay width of $f_0(980)$ is shifted from $0.207$ GeV obtained from GLSM to $0.070$ GeV in our model which is closer to the experimental value. In Fig \ref{pipiscat}, we have compared the predictions of our model for the real part of the K-matrix unitarized $\pi\pi$ scattering amplitude for scenario II (for the minimum achieved $\chi$ which equals $1.19$)  with the results of GLSM for three different choices of  $m[\pi(1300)]$ and $A_3/A_1=20,\, 30$  \cite{40soode2,soode2}.
It can be seen that the prediction of GLSM for the K-matrix unitarized $\pi \pi$ scattering amplitude for $m[\pi(1300)]=1.22$ and  $A_3/A_1=30$  is  in better agreement with experiment below $1$ GeV, while adding scalar glueball to SNLSM (the solid line in Fig. \ref{pipiscat}) gives a better fit with the experimental data above $1$ GeV. Also shown in Fig. \ref{pipiscatmean} are the averages (triangles) and standard deviations (error bars) of  the prediction of GLSM for the real part of the K-matrix unitarized $\pi\pi$ scattering amplitude resulted from variation of $m[\pi(1300)]$ and $A_3/A_1$ . Comparing this with the average values (squares) together with standard deviations (error bars) of SNLSM with glueball  (stemming from averaging over all the sets for which $\chi<\chi_{\rm exp.}$), shows that up to about $1.1$ GeV, the results coincide, although in some regions matching with experiment (dots with error bars) is not seen.
 From Fig. \ref{pipiscatmean} it can also be seen that while the predictions of SNLSM with glueball for the region of $1.1-1.7$ GeV, do match with observed data, GLSM misses agreement with data. \\
 In order to illustrate that GLSM does not succeed in predicting the behavior of data for regions above $1$ GeV, it is worthwhile to also compare its prediction for $\pi\pi$ scattering phase shift with those of SNLSM with and without glueball. From Fig. \ref{pipiscatph}, it is evident that while up to about $1$ GeV, the prediction of SNLSM without glueball match well with experiment, the prediction of SNLSM including glueball (scenario II) follows the behavior of data up to about $1.5$ GeV and the agreement with data is rather good for $\sqrt{s}=1.1-1.5$ GeV. It is clear that for the energy region below $1.1$ GeV, the predictions of GLSM and our model are close and far from experiment but for $\sqrt{s}>1.1$ GeV, our model is consistent with observed data. Also it can be seen that the predictions of LO ChPT \cite{ollercua,pelaezrev,pelaezrev} with different unitarization approaches are successful up to values of $\sqrt{s}=1.2$ GeV and from there on a fair description of data is achieved.  Note that at this stage it does not make sense to compare the predictions of our model and \cite{ollernd,ollercua} since the coupled channels are absent in our calculations.

\begin{table}[!htbp]
	\footnotesize
	\caption{The estimate of SNLSM in the absence and  presence of the scalar glueball for scenario II.}
	\resizebox{1\hsize}{!}{
				
		\begin{tabular}{@{}|c|cc|cc|cc|@{}}
				\noalign{\hrule height 1pt}
				\noalign{\hrule height 1pt}
				& \multicolumn{2}{c|}{\textbf{SNLSM without glueball}}& \multicolumn{2}{c|}{\textbf{SNLSM with glueball (scenario II)}}& \multicolumn{2}{c|}{\textbf{Experimental values}} \\
				\noalign{\hrule height 1pt}
				& Width (GeV) & Mass  of decaying & Width (GeV)  & Mass of decaying  & Width (GeV)  & Mass of decaying  \\    
				&  &  particle (GeV)& & particle (GeV) & & particle  (GeV) \\
				\noalign{\hrule height 1pt}
				
				$\sigma $  &  $0.562 \pm 0.022 $ & $0.454 \pm  0.001 $&  $0.350 \pm 0.009$ & $0.460 \pm 0.001$ & $0.400\, \rm{to}\, 0.700$ &   $0.400\, \rm{to}\, 0.550$  \\
				$f_0(980)$  & $ 2.310\times 10^{-6}\pm  5.505\times 10^{-7} $ & $1.362 \pm 0.229 $ &  $ 0.070 \pm 0.001$ & $ 1.066 \pm 0.006$  & $0.040\, \rm{to}\, 0.100$ &   $0.990 \pm 0.020$   \\
				$f_3  $ & ..&... &  $0.077 \pm 0.013$ & $1.566 \pm 0.046$&$0.109 \pm .007$&$1.505 \pm .006$    \\
				$K_0^*(800) $ or $\kappa $  & $0.524 \pm 0.020$ & $0.796\pm 0.007$  &  $0.451\pm  0.001$ & $0.765 \pm 0.004$& $0.547 \pm 0.024$  &  $0.682 \pm 0.029 $    \\
				$a_0(980) $  &  $ 0.150 \pm   0.028$ & $0.867 \pm 0.017 $&  $0.070 \pm 009$ & $0.777 \pm 0.003$& $0.050 \, \rm{to}\, 0.100$ & $0.980 \pm 0.020$   \\
				\noalign{\hrule height 1pt}
				\noalign{\hrule height 1pt}
			\end{tabular}\label{table_conclusion}}
	\end{table}
	
\begin{table}[!htbp]
	\small
	\centering
	\caption{Predicted physical masses and decay widths of the lowest lying mesons in the absence of glueball  in GLSM \cite{40soode2,soode2} and in the presence  of it in SNLSM obtained from the unitarized amplitudes of $\pi \pi$, $\pi K$ and $\pi \eta$ scatterings.}
	{\begin{tabular}{@{}|c|cc|cc|@{}}
			\noalign{\hrule height 1pt}
			\noalign{\hrule height 1pt}
				& \multicolumn{2}{c!{\vrule width .5pt}}{\textbf{GLSM\cite{40soode2,soode2}}} &  \multicolumn{2}{c!{\vrule width .5pt}}{\textbf{{SNLSM with glueball}}} \\
			& Width (GeV)  & Mass (GeV) &   Width (GeV)  & Mass (GeV)  \\
			\noalign{\hrule height 1pt}
			$\sigma $   &  $0.385 \pm 0.061$ & $0.476 \pm 0.004$             &    $0.350 \pm 0.009$ & $0.460 \pm 0.001$    \\
			$f_0(980)$  &    $0.207 \pm 0.065$ & $1.053 \pm 0.044$                 & $ 0.070 \pm 0.001$& $ 1.066 \pm 0.006$   \\
			$\kappa $   &    $0.689 \pm 0.027$ & $0.722 \pm 0.028$   & $0.451\pm  0.001$ & $0.765 \pm 0.004$ \\
			$a_0(980) $ &    $0.060 \pm 0.052$  & $0.984 \pm 0.007$  &  $0.070 \pm 009$ & $0.777 \pm 0.003$ \\
			\noalign{\hrule height 1pt}
			\noalign{\hrule height 1pt}
		\end{tabular}\label{two_body2}}
\end{table}

\par
The model presented in this paper clearly is more successful in predicting the masses and decay widths of the scalar and pseudoscalar mesons and also the real part of the unitarized  $\pi\pi$ scattering amplitude in comparison with the standard SNLSM and also GLSM in some cases. However, it misses next-to-lowest lying scalar and pseudoscalar mesons and also ignores mixing among two quark and four quark states and as a consequence cannot estimate four-quark percentages of isoscalars.  Furthermore, if we calculate gloun condensate from Eq. (\ref{scaleanomaly}), we obtain $C=0.247$ GeV which does not agree with the results of QCD  sum-rules or lattice simulations. Using Eq. (\ref{lamda}), the value of $\Lambda=0.483$ GeV is evaluated which is far from the expected value in PDG ($332 \pm 17$ MeV) \cite{pdg}.
\par

 In view of the results presented in \cite{soode2}, the GLSM cannot estimate  reasonable widths for some scalars above $1$ GeV \cite{soode2}. We hope that adding the scalar glueball to GLSM will have a considerable effect on the predictions.
Therefore, the next step of this line of research would be adding scalar glueball to the GLSM Lagrangian. 

\section*{Acknowledgements}
We would like to show our gratitude to A.H. Fariborz without whom this work would not have been possible. S.M.Z. wish to thank Shiraz University research council. S.Z. appreciate the support of Sistan and Baluchestan University research council.

\newpage
\appendix
\section{A Brief review on Generalized Linear Sigma Model}\label{appGLSM}

The basic feature of Generalized Linear Sigma Model is considering mixing between two chiral nonets (a two quark nonet, and a four quark nonet) below 2 GeV \cite{f_global,f_chiral}. The Lagrangian of the model which contains two scalar meson nonets and two pseudoscalar meson nonets is defined as

\begin{equation}
{\cal L} = - \frac{1}{2} {\rm Tr}
\left( \partial_\mu M \partial_\mu M^\dagger
\right) - \frac{1}{2} {\rm Tr}
\left( \partial_\mu M^\prime \partial_\mu M^{\prime \dagger} \right)
- V_0 \left( M, M^\prime \right) - V_{SB},
\label{mixingLsMLag}
\end{equation}
where $M$ and $M'$ are 3$\times$3 matrix
chiral nonet fields. $M$ represents the ``bare"
quark-antiquark scalar and pseudoscalar nonet fields while
$M'$ describes ``bare" scalar and pseudoscalar fields containing two quarks and two antiquarks.
\begin{equation}
M = S +i\phi, \hskip 2cm
M^\prime = S^\prime +i\phi^\prime.
\label{sandphi1}
\end{equation}
   The potential of the model $V_0(M,M^\prime)$ is defined in terms of $M$ and $M'$ and based on the number of underlying quark and antiquark lines in each term in the potential, $N$, a criterion for limiting the number of terms at each order of calculation. Up to $N=8$, the potential reads
   
   \begin{eqnarray}
   V_0 =&-&c_2 \, {\rm Tr} (MM^{\dagger}) +
   c_4^a \, {\rm Tr} (MM^{\dagger}MM^{\dagger})
   \nonumber \\
   &+& d_2 \,
   {\rm Tr} (M^{\prime}M^{\prime\dagger})
        + e_3^a(\epsilon_{abc}\epsilon^{def}M^a_dM^b_eM'^c_f + {\rm H. c.})
   \nonumber \\
        &+&  c_3\left[ \gamma_1 {\rm ln} (\frac{{\rm det} M}{{\rm det}
   M^{\dagger}})
   +(1-\gamma_1){\rm ln}\frac{{\rm Tr}(MM'^\dagger)}{{\rm
   Tr}(M'M^\dagger)}\right]^2,
   \label{SpecLag}
   \end{eqnarray}
  
where all the terms are SU(3)$_{\rm L} \times$ SU(3)$_{\rm R}$ (but not necessarily U(1)$_{\rm A}$) invariants. The possible symmetry breaking term $V_{SB}$ which models the QCD mass term is
\begin{equation}
V_{SB} = - 2\, {\rm Tr} (A\, S),
\label{vsb}
\end{equation}
where $A={\rm diag} (A_1,A_2,A_3)$ are proportional to the three light quark
current masses (i.e., in the isospin invariant limit $A_1=A_2 \propto m_u=m_d$ and $A_3\propto m_s$.)
Up to this order, the model has twelve unknown parameters: the six coupling constants
 given in Eq. (\ref{SpecLag}), the two quark mass parameters,
($A_1=A_2,A_3$) and the four vacuum parameters ($\alpha_1
=\alpha_2=\langle S_1^1 \rangle,\alpha_3=\langle S_3^3 \rangle,\beta_1=\beta_2=\langle {S'}_1^1 \rangle,\beta_3=\langle {S'}_3^3 \rangle$). These parameters are determined via four minimum
equations and eight experimental inputs.
  \par
The model has been used to explore the underlying mixings among scalar mesons below and above 1 GeV (as well as those of their pseudoscalar chiral partners). Up to now, the underlying mixings among scalar mesons in the $\eta' \rightarrow \eta \pi \pi$ decay \cite{etapetapipi} and also in $\pi \pi$ \cite{40soode2}, $\pi K$ \cite{41soode2} and $\pi \eta$ \cite{pieta} scatterings are investigated exploiting this model. It is found that while the single nonet linear sigma model which only includes lowest-lying nonet is not accurate in predicting the decay widths and the amplitudes, but taking into account the mixing of this nonet with the next-to-lowest-lying nonet, and also considering the effect of the final-state interactions, significantly improves the results. This confirms the global picture of scalar mesons: those below 1 GeV are predominantly four-quark states and those above 1 GeV, are closer to the conventional p-wave quark-antiquark states. 
\par
Despite being successful in predicting the properties of lowest-lying scalars and pseudoscalars and also the scattering amplitudes in region up to about $1$ GeV, the model predictions for the widths and masses of next-to-lowest-lying scalar and pseudoscalar mesons besides the scattering amplitudes above $1$ GeV are not close to experiment. This encourages us to further improve the Lagrangian of the model to also includes the terms of mixing among scalars and glueballs and we believe that it will considerably enhances the results.

\section{K-matrix unitarized amplitudes of $\pi \pi$, $\pi K$ and $\pi \eta$ scatterings.}\label{appA}

\subsection{$\pi \pi$ scattering}

The $I=J=0$ bare partial wave scattering amplitude of $\pi \pi$ scattering consists of a constant background and two or three poles corresponding to the two lowest-lying isosinglet scalars ($\sigma$ and $f_0(980)$) and $f_3$ (which may be one of the next to lowest lying isoscalar scalars in the case of adding glueball to the Lagrangian)

\begin{equation}
	{T_0^0}^B = T_\alpha + \sum_i^{n_f}  {  {T_\beta^i } \over
		{m_{f_i}^2 - s}},
	\label{T00B}
\end{equation}
with
\begin{eqnarray}
	T_\alpha &=&
	{1\over 64 \pi}
	\sqrt{1 - {4 m_\pi^2\over s}}\,
	\left[-5\, \gamma^{(4)}_{\pi \pi} +
	{ 2 \over {p_\pi^2}}\,
	\sum_i^{n_f} \gamma_{f_i\pi \pi}^2\,  {\rm ln} \left(1 +  {{4
			p_\pi^2}\over m_{f_i}^2} \right)
	\right],
	\nonumber \\
	T_\beta^i &=&
	{3\over 16 \pi}
	\sqrt{1 - {4 m_\pi^2\over s}}\, \gamma_{f_i\pi \pi}^2,
	\label{pipiformula}
\end{eqnarray}
where $p_\pi = \sqrt{s - 4 m_\pi^2} / 2$ and $n_f$ is two (three) for single nonet without (with) glueball . The three and four point couplings, i.e., $\gamma_{f_i\pi \pi}$ and $\gamma^{(4)}_{\pi \pi}$,  are defined by the Lagrangian density 
\begin{eqnarray}
	-{\cal L} &=& \gamma^{(4)}_{\pi \pi} (\mbox{\boldmath ${\pi}$} \cdot{\mbox{\boldmath ${\pi}$}})^2+ \gamma^{(4)}_{\pi K}\overline{K} K \mbox{\boldmath ${\pi}$} \cdot{\mbox{\boldmath ${\pi}$}}+\gamma^{(4)}_{\pi \eta}\eta \eta \, \mbox{\boldmath ${\pi}$} \cdot{\mbox{\boldmath ${\pi}$}}\nonumber\\
	&&+\frac{\gamma_{f_i \pi \pi}}{\sqrt 2}
	f_i \mbox{\boldmath ${\pi}$} \cdot{\mbox{\boldmath ${\pi}$}}+ \frac{\gamma_{f_i K K}}{\sqrt 2}f_i K\overline{K}
	+ \frac{\gamma_{a KK}}{\sqrt 2} \overline{K} {\mbox{\boldmath ${\tau}$}}  \cdot {\bf a} K+ \frac {\gamma_{\kappa K \pi}}{\sqrt 2}(\overline{K}{\mbox{\boldmath ${\tau}$}} \cdot {\mbox{\boldmath ${\pi}$}} \kappa+\rm{H.c.})\nonumber\\
	&&+\,\gamma_{\kappa {K} \eta} \left({\overline{ \kappa}}  K  {\eta} +\rm{H.c.} \right)+\gamma_{\kappa {K} \eta '} \left(
	{\bar \kappa}  K  {\eta '} + \rm{H.c.} \right)\,+ \gamma_{a \pi\eta} {\bf a} \cdot  \mbox{\boldmath
		${\pi}$} \eta+ \gamma_{a \pi\eta'} {\bf a} \cdot  \mbox{\boldmath${\pi}$}  \eta'  \nonumber \\
	&&+\, \gamma_{f_i \eta \eta} f_i  \eta  \eta+ \gamma_{f_i \eta \eta'} f_i  \eta \eta'+ \gamma_{f_i\eta' \eta'} f_i  \eta'\eta' + \cdots,
	\label{lagiso}
\end{eqnarray}
where the subscript $ i (=1,2$ and $ 3 )$ shows the different isosinglet meson states. These isomultiplets contain the physical fields
\begin{eqnarray}\label{isomul}
	&&K=
	\left[
	\begin{array}{cc}
		K^+\\
		K^0
	\end{array}
	\right],\qquad
	\overline{K}=
	\left[
	K^-\; \overline{K}^0
	\right],\qquad
	\kappa=
	\left[
	\begin{array}{cc}
		\kappa^+\\
		\kappa^0
	\end{array}
	\right],\qquad
	\overline{\kappa}=
	\left[
	\kappa^-\; \overline{\kappa}^0
	\right],\nonumber\\
	&&\pi_1=\frac{1}{\sqrt{2}}(\pi^+ +\pi^-),\qquad \pi_2=\frac{i}{\sqrt{2}}(\pi^+ - \pi^-),\qquad \pi_3=\pi^0, \nonumber\\
	&&a_{01}=\frac{1}{\sqrt{2}}(a_0^+ +a_0^-),\qquad a_{02}=\frac{i}{\sqrt{2}}(a_0^+ - a_0^-),\qquad a_{03}=a_0^0.
\end{eqnarray}
Making use of Eq. (\ref{lagiso}) and differentiating with respect to appropriate fields, the three and four point couplings are related to bare couplings 
\begin{equation}
	\gamma^{(4)}_{\pi \pi}= \left\langle \frac{\partial^4 V}{\partial \phi_1^2 \partial\phi_2^1\partial\phi_1^2\partial\phi_2^1}\right\rangle_0,
\end{equation}
and
\begin{equation}\label{L00}
	\gamma_{f_i \pi \pi}={1\over \sqrt{2}}\,\sum_A \left\langle \frac{\partial^3 V}{\partial f_A \partial\phi_1^2 \partial\phi_2^1}\right\rangle_0 (L_0)_{Ai},
\end{equation}
where $A$ ia a placeholder for $a$, $b$, and $c$, that, respectively, represent the three bases in Eq. (\ref{basis}) 

\begin{equation}\label{basis}
F_{0}=   \begin{bmatrix}
f_a \\
f_b \\
f_c
\end{bmatrix}=\begin{bmatrix}
\dfrac{S_1^1+S_2^2}{\sqrt{2}} \propto n \bar{n} \\
S_3^3\propto s\bar{s} \\
\alpha_{s}G_{\mu\nu}G^{\mu\nu}
\end{bmatrix},
\end{equation}
where $n$ and  $s$ respectively denote the non-strange and strange quark content and $G^{\mu\nu}$ is the field-strength tensor of gluon fields. 
$L_0$ in Eq. (\ref{L00})  is the rotation matrix describing the underlying mixing among two (three) isoscalar fields
\begin{equation}
 \begin{bmatrix}
f_1   \\
f_2    \\
f_3
\end{bmatrix}=L_{0}^{-1}F_{0},
\end{equation}
where $f_1$ and $f_2$ are clearly identified with $f_0(500)$ and $f_0(980)$ respectively, and $f_3$ resembles one of the three isoscalar scalars above 1 GeV, i.e., $f_0(1370)$, $f_0(1500)$ and $f_0(1710)$ which may represent the scalar glueball. $F_0$  contains the non-physical fields 
 The bare couplings are given in Appendix \ref{appB}.

In order to consider final state interactions, avoiding the divergence of the bare amplitude at resonance masses and also forcing unitarity of S-matrix at all $s$ for the partial wave amplitude of $\pi \pi$ scattering, the K-matrix unitarization method \cite{40soode2,41soode2,pieta} which was originally introduced by Wigner \cite{wigner} may be applied. In this method,  the partial wave bare amplitude $T^{I\, B}_{l}$ transforms to unitarized amplitude $T^I_l$ through the following equation
\begin{equation}\label{TIL}
	T^I_l=\frac{T^{I\,B}_{l}}{1-i T^{I\, B}_{l}},
\end{equation}
where $I$ and $l$ are the partial wave isospin and angular momentum.  The physical masses $(\tilde{m}_i)$ and full decay widths $(\tilde{\Gamma}_i)$ of the intermediate scalar mesons are found from the poles $(z_i)$ of the K-matrix unitarized amplitude  

\begin{equation}\label{poles}
	1- i T_l^{I\,B}=0\Longrightarrow z_i=\tilde{m}_i^2- i\tilde{m}_i \tilde{\Gamma}_i.
\end{equation}

With the potential of Eq. (\ref{withotglue}) and using Eq. (\ref{T00B}), we will find the $\pi \pi$ scattering bare amplitude. Inserting that in Eq. (\ref{poles}), the physical masses and widths of $f_i$ mesons can be found. Likewise, the physical masses and widths of $\kappa$ and $a_0(980)$ mesons are obtained from the roots of the denominator of (\ref{TIL}) for $I=1/2, J=0$ channel of $\pi K$ and $I=1, J=0$ channel of $\pi \eta$ scatterings respectively.

\subsection{$\pi K$ scattering}
The $I=1/2$ $\pi K$ tree level amplitude involves $\kappa$  exchange in the $s$ and $u$ channels, $f_i$ exchanges in the
$t$ channel as well as a four point contact term. The 
tree level invariant amplitude may be written as

\begin{eqnarray}\label{a}
A^{\frac{1}{2}}(s,t,u)= -\gamma_{\pi K}^{(4)}+\frac{3}{2} \frac{\gamma_{\kappa  \pi K }^2}{m_{\kappa_i}^2-s}-\frac{1}{2} \frac{\gamma_{\kappa\pi K}^2}{m_{\kappa_i}^2-u}+\sum_{i=1}^{n_f} \frac{\gamma_{f_{j}KK}\gamma_{f_{i}\pi \pi}}{m_{f_i}^2-t}.
\end{eqnarray}
The couplings are defined by the Lagrangian density in Eq.(\ref{lagiso}) and are related to bare couplings by
\begin{eqnarray}
\gamma^{(4)}_{\pi K} &=& \left\langle \frac{\partial^4 V}{\partial \phi_1^2 \partial\phi_2^1\partial\phi_1^3\partial\phi_3^1}\right\rangle_0,\nonumber\\
\gamma_{\kappa \pi K}&=&\sum_a \left\langle \frac{\partial^3 V}{\partial\phi_1^2 \partial S_2^3 \partial\phi_3^1 }\right\rangle_0,  \nonumber\\
\gamma_{f_i K K}&=& \sqrt{2} \,\sum_A \left\langle \frac{\partial^3 V}{\partial f_A \partial\phi_1^3 \partial\phi_3^1}\right\rangle_0 (L_0)_{Ai}.
\end{eqnarray}
The $J=0$ partial wave amplitude can be found from
\begin{eqnarray}\label{partial}
T_0^{\frac{1}{2}B}=\frac{\rho(s)}{2}\int_{-1}^1 d \cos\theta P_0(\cos \theta)A^\frac{1}{2}(s,t,u),
\end{eqnarray}
with $\rho(s)=q /(8\pi\sqrt{s})$ where $q$ is the center of mass momentum $q=1/(2\sqrt{s})\sqrt{(s-(m_{\pi}+m_K)^2)(s-(m_{\pi}-m_K)^2)}$.
Performing the partial wave projection we find the ``bare'' $I=1/2$, $J=0$ amplitude
\begin{eqnarray}
T_0^{\frac{1}{2}B}&=&\frac{\rho (s)}{2}\left[-2\gamma_{\pi K}^{(4)}+3 \frac{\gamma_{\kappa \pi K}^2}{m_{\kappa}^2-s}
-\frac{1}{4q^2} {\gamma_{\kappa\pi K}^2}\ln{\left(\frac{B_{\kappa}+1}{B_{\kappa}-1}\right)}
+\frac{1}{2q^2}\sum_{i=1}^{n_f} \gamma_{f_{j}KK}\gamma_{f_{i}\pi \pi}\ln{\left(1+\frac{4q^2}{m_{f_i}^2}\right)}\right],
\label{T012_B}
\end{eqnarray}
in which
\begin{eqnarray}
B_{\kappa}=\frac{1}{2q^2}\left[(m_{\kappa})^2-m_K^2-m_{\pi}^2+2\sqrt{(m_{\pi}^2+q^2)(m_K^2+q^2)}\right],
\end{eqnarray}
and the Mandelstam variables are expressed in terms of $q$ and $\theta$
\begin{eqnarray}
t &=& 2 m_{\pi}^2-2 (q^2+m_{\pi}^2)+2 q^2 \cos \theta,
\nonumber \\
u &=&  m_{\pi}^2+m_K^2+2\sqrt{(m_{\pi}^2+q^2)(m_K^2+q^2)}-2q^2 \cos{\theta}.
\end{eqnarray}
Unitarizing the bare amplitude of (\ref{T012_B}) via K-matrix method, the physical mass and width of $\kappa$ resonance will be predicted.
\subsection{$\pi \eta$ scattering}
To this end, the tree level $I=1$  $\pi \eta$ invariant amplitude is given by
\begin{equation}
A(s,t,u)=-\gamma_{\pi \eta}^{(4)} +
\sum _{i=1} ^{n_f}\frac{{{2 \sqrt{2}}}\,\gamma_{f_i\pi \pi} \gamma_{f_i\eta \eta}}{m_{f_i}^2-t}+ \gamma_{a \pi \eta}^2\left[\frac{1}{m_{a}^2-s}+\frac{1}{m_{a}^2-u}\right],
\label{Astu_template}
\end{equation}
where the coupling constants are defined as
\begin{eqnarray}
\gamma^{(4)}_{\pi\eta}&=&\sum_{A,B} \left\langle \frac{\partial^4 V}{\partial \phi_1^2 \partial\phi_2^1\partial\eta_A\partial\eta_B}\right\rangle_0 (R_{0})_{A1}(R_{0})_{B1},\nonumber\\
\gamma_{f_i\eta\eta}&=&\frac{1}{2}\sum_{A,B,C}\left\langle \frac{\partial^3 V}{\partial f_A \partial \eta_B \partial \eta_C}\right\rangle _0 (L_0)_{A1} (R_0)_{B1} (R_0)_{C1},\nonumber\\
\gamma_{a_0 \pi\eta}&=&\sum_A \left\langle \frac{\partial^3 V}{\partial S_1^2 \partial\eta_A\partial\phi_2^1}\right\rangle_0 (R_{0})_{A1},\nonumber\\
\end{eqnarray}
where $R_0$ is the $I=0$ pseudoscalars rotation matrix 
\begin{equation}
\begin{bmatrix}
\eta_{1}   \\
\eta_{2}   
\end{bmatrix}=R_{0}^{-1}\begin{bmatrix}
\eta_{a}\\
\eta_{b}   
\end{bmatrix},
\end{equation}
and $\eta_a=(\phi_1^1+\phi_2^2)/\sqrt{2}$ and $\eta_b=\phi_3^3$.

The ``bare'' $J=0$ partial wave amplitude (s-wave) is obtained from Eq. (\ref{partial})
\begin{eqnarray}\label{pietatem}
T_{0}^{1\,B}= \frac{q(s)}{16 \pi \sqrt{s}}\Bigg[&&- 2 \gamma_{\pi \eta}^{(4)} + \gamma_{a \pi \eta}^2\left( \frac{1}{2q^2}\ln \left(\frac{(B_{\eta})+1}{(B_{\eta})-1}\right)+\frac{2}{m_{a}^2-s}\right)\nonumber\\
&&+\sum_{i=1}^{n_f} \frac{{{\sqrt{2}}}}{q^2} \gamma_{f_i \eta \eta} \gamma_{f_i \pi \pi}\ln \left(1+\frac{4 q^2}{m_{f_i}^2}\right)\Bigg],
\end{eqnarray}
where q is the center of mass momentum 
\begin{equation}
q= \frac{1}{2\sqrt{s}}\sqrt{(s-(m_{\pi}+m_{\eta})^2)(s-(m_{\pi}-m_{\eta})^2)},
\end{equation}
and $B_{\eta}$ is defined as
\begin{equation}
B_{\eta} = \frac{1}{2q^2}\left[m_{a}^2-m_{\pi}^2-m_{\eta}^2+2 \sqrt{(m_{\pi}^2+q^2)(m_{\eta}^2+q^2)}\right].
\end{equation}
Here the Mandelstam variables are 
\begin{eqnarray}
t &=& -2q^2(1-\cos \theta) \nonumber \\
u &=& m_{\eta}^2 + m_{\pi}^2 - 2 \sqrt{(m_{\pi}^2 + q^2)(m_{\eta}^2 + q^2)} -2 q^2 \cos \theta,
\end{eqnarray}
where $\theta$ is the scattering angle. As before, we get the physical properties of $a_0(980)$ by unitarizing the bare amplitude of Eq. (\ref{pietatem}) and solving for the roots of the denominator of Eq. (\ref{TIL}).

\section{Bare three- and four-point coupling constants}\label{appB}
\subsection{SNLSM without glueball}

\begin{eqnarray}\label{vertexf2pipi}
\left\langle  \frac{\partial^4 V}{\partial \phi_1^2 \partial \phi_2^1 \partial \phi_1^2 \partial \phi_2^1}\right\rangle  &=& 8 \Big(c_4^a+2\, c_4^b+3(4\,c_6^b+c_6^a)\alpha_1^2+6\, c_6^b\alpha_3^2)\\
\left\langle  \frac{\partial^4 V}{\partial \phi_1^2 \partial \phi_2^1 \partial \phi_1^3 \partial \phi_3^1}\right\rangle  &=& 4\left( {c_4^a + 2c_4^b} \right) + 12\left( {4c_6^b + {c_6^a}} \right)\alpha _1^2 - 6{c_6^a}{\alpha _1}{\alpha _3} + 6\left( {4c_6^b + {c_6^a}} \right)\alpha _3^2\\
\left\langle  \frac{\partial^4 V}{\partial \phi_1^2 \partial \phi_2^1 \partial \eta_a \partial \eta_a}\right\rangle  &=&  4\left( {3c_4^a + 2c_4^b + \frac{{8{c_3}}}{{\alpha _1^4}} + 3\left( {4c_6^b + 3{c_6^a}} \right)\alpha _1^2 + 6c_6^b\alpha _3^2} \right)\\
\left\langle  \frac{\partial^4 V}{\partial \phi_1^2 \partial \phi_2^1 \partial \eta_a \partial \eta_b}\right\rangle  &=&  \frac{{8\sqrt 2 {c_3}}}{{\alpha _1^3{\alpha _3}}}\\
\left\langle  \frac{\partial^4 V}{\partial \phi_1^2 \partial \phi_2^1 \partial \eta_b \partial \eta_b}\right\rangle  &=&  8\left( {c_4^b + 6c_6^b\alpha _1^2 + 3c_6^b\alpha _3^2} \right)\\
\left\langle  \frac{\partial^3 V}{\partial f_a \partial \phi_1^2 \partial \phi_2^1}\right\rangle  &=& 4 \sqrt{2} \alpha_1\Big(c_4^a+2\,c_4^b+3(4\,c_6^b+c_6^a)\alpha_1^2+6\,c_6^b\alpha_3^2\Big)\\
\left\langle  \frac{\partial^3 V}{\partial f_b \partial \phi_1^2 \partial \phi_2^1}\right\rangle  &=& 8 \alpha_3 \Big( c_4^b + 6\, c_6^b  \alpha_1^2 + 3\, c_6^b \alpha_3^2\Big)\\
\left\langle  \frac{\partial^3 V}{\partial f_a \partial \phi_1^3 \partial \phi_3^1}\right\rangle  &=&
\sqrt 2 \,\Bigg(12\left( {4\,c_6^b + c_6^a} \right)\alpha _1^3 - 2\,c_4^a{\alpha _3} - 9\,c_6^a\alpha _1^2{\alpha _3} - 3\,c_6^a\alpha _3^3\nonumber\\
&&\quad \; \; + {\alpha _1}\Big( {4\left( {c_4^a + 2\,c_4^b} \right) + 6\left( {4\,c_6^b + c_6^a} \right)\alpha _3^2} \Big)\Bigg)
\\
\left\langle  \frac{\partial^3 V}{\partial f_b \partial \phi_1^3 \partial \phi_3^1}\right\rangle  &=& 
2\,\Bigg( - 3\,c_6^a\alpha _1^3 + 6\left( {4\,c_6^b + c_6^a} \right)\alpha _1^2{\alpha _3} - {\alpha _1}\left( {2\,c_4^a + 9\,c_6^a\alpha _3^2} \right)\nonumber\\
&&\quad  + 4\,{\alpha _3}\Big( {c_4^a + c_4^b + 3\left( {c_6^b + c_6^a} \right)\alpha _3^2} \Big)\Bigg)
\\
\left\langle  \frac{\partial^3 V}{\partial \phi_3^1 \partial \phi_1^2 \partial s_2^3}\right\rangle  &=& 2{\alpha _3}\left( {2c_4^a + 3{c_6^a}\alpha _1^2 + 3{c_6^a}\alpha _3^2} \right)\\
\left\langle  \frac{\partial^3 V}{\partial \eta_a \partial S_1^2 \partial \phi_2^1 }\right\rangle  &=&\frac{{4\sqrt 2 \left( {2{c_3} + c_4^a\alpha _1^4 + 3{c_6^a}\alpha _1^6} \right)}}{{\alpha _1^3}}\\
\left\langle  \frac{\partial^3 V}{\partial \eta_b \partial S_1^2 \partial \phi_2^1 }\right\rangle  &=&\frac{{8{c_3}}}{{\alpha _1^2{\alpha _3}}}\\
\left\langle  \frac{\partial^3 V}{\partial f_a \partial \eta_a \partial \eta_a }\right\rangle  &=& \frac{{4\sqrt 2 \Big( {4{c_3} + 3\left( {4c_6^b + {c_6^a}} \right)\alpha _1^6 + \alpha _1^4\left( {c_4^a + 2c_4^b + 6c_6^b\alpha _3^2} \right)} \Big)}}{{\alpha _1^3}}\\
\left\langle  \frac{\partial^3 V}{\partial f_a \partial \eta_a \partial \eta_b }\right\rangle  &=& \frac{{8{c_3}}}{{\alpha _1^2{\alpha _3}}}\\
\left\langle  \frac{\partial^3 V}{\partial f_a \partial \eta_b \partial \eta_b }\right\rangle  &=& 8\sqrt 2 {\alpha _1}\left( {c_4^b + 6c_6^b\alpha _1^2 + 3c_6^b\alpha _3^2} \right)\\
\left\langle  \frac{\partial^3 V}{\partial f_b \partial \eta_a \partial \eta_a }\right\rangle  &=& 8{\alpha _3}\left( {c_4^b + 6c_6^b\alpha _1^2 + 3c_6^b\alpha _3^2} \right)\\
\left\langle  \frac{\partial^3 V}{\partial f_b \partial \eta_b \partial \eta_a }\right\rangle  &=& \frac{{8\sqrt 2 {c_3}}}{{{\alpha _1}\alpha _3^2}}\\
\left\langle  \frac{\partial^3 V}{\partial f_b \partial \eta_b \partial \eta_b }\right\rangle  &=& \frac{{16{c_3}}}{{\alpha _3^3}} + 8\left( {c_4^a + c_4^b + 6c_6^b\alpha _1^2} \right){\alpha _3} + 24\left( {c_6^b + {c_6^a}} \right)\alpha _3^3
\end{eqnarray}

\subsection{SNLSM with glueball}

\begin{eqnarray}
\left\langle  \frac{\partial^3 V}{\partial \phi_1^2 \partial \phi_2^1 \partial \phi_1^2 \partial \phi_2^1}\right\rangle  &=&
8\,  c_4^a + 16 \, c_4^b + \frac{24 \, \alpha_1^2  c_6^a  \gamma^2}{h_0^2} - \frac{2\,  c'  h_0^4}{3\,  \alpha_1^4  \gamma^4} + \frac{48\,  c_6^b  \gamma^2  (2\,  \alpha_1^2 + \alpha_3^2)}{h_0^2}\\
\left\langle  \frac{\partial^4 V}{\partial \phi_1^2 \partial \phi_2^1 \partial \phi_1^3 \partial \phi_3^1}\right\rangle  &=& 
4\,c_4^a + 8\,c_4^b + \frac{{18\,{\gamma ^6}\alpha _1^3{\alpha _3}\Big( {8\,c_6^b\alpha _1^2 + 16\,{c^2}\alpha _3^2 + c_6^a\left( {2\,\alpha _1^2 - {\alpha _1}{\alpha _3} + \alpha _3^2} \right)} \Big) - h_0^6c'}}{{3\,{\gamma ^4}h_0^2\alpha _1^3{\alpha _3}}}\\
\left\langle  \frac{\partial^4 V}{\partial \phi_1^2 \partial \phi_2^1 \partial \eta_a \partial \eta_a}\right\rangle  &=&
12\,c_4^a + 8\,c_4^b + \frac{{36\,\alpha _1^2c_6^a{\gamma ^2}}}{{h_0^2}} + \frac{{32\,{c_3}h_0^4}}{{\alpha _1^4{\gamma ^4}}} - \frac{{c'h_0^4}}{{\alpha _1^4{\gamma ^4}}} + \frac{{24\,c_6^b{\gamma ^2}\left( {2\alpha _1^2 + \alpha _3^2} \right)}}{{h_0^2}}\\
\left\langle  \frac{\partial^4 V}{\partial \phi_1^2 \partial \phi_2^1 \partial \eta_a \partial \eta_b}\right\rangle  &=&
\frac{{8\sqrt 2 \,{c_3}h_0^4}}{{\alpha _1^3{\alpha _3}{\gamma ^4}}}\\
\left\langle  \frac{\partial^4 V}{\partial \phi_1^2 \partial \phi_2^1 \partial \eta_b \partial \eta_b}\right\rangle  &=&
8\,c_4^b + \frac{{24\,c_6^b{\gamma ^2}\left( {2\,\alpha _1^2 + \alpha _3^2} \right)}}{{h_0^2}}\\
\left\langle  \frac{\partial^3 V}{\partial f_a \partial \phi_1^2 \partial \phi_2^1}\right\rangle  &=&
\sqrt 2 \left( {4\,{\alpha _1}c_4^a + 8\,{\alpha _1}c_4^b + \frac{{12\,\alpha _1^3 c_6^a{\gamma ^2}}}{{h_0^2}} - \frac{{c'h_0^4}}{{3\, \alpha _1^3{\gamma ^4}}} + \frac{{24\,{\alpha _1}c_6^b{\gamma ^2}\left( {2\,\alpha _1^2 + \alpha _3^2} \right)}}{{h_0^2}}} \right)\\
\left\langle  \frac{\partial^3 V}{\partial f_b \partial \phi_1^2 \partial \phi_2^1}\right\rangle  &=&
8\,{\alpha _3}c_4^b + \frac{{24\,{\alpha _3}c_6^b{\gamma ^2}\left( {2\,\alpha _1^2 + \alpha _3^2} \right)}}{{h_0^2}}\\
\left\langle  \frac{\partial^3 V}{\partial f_c \partial \phi_1^2 \partial \phi_2^1}\right\rangle  &=&
\frac{{4\,{c_2}{h_0}}}{{{\gamma ^2}}} - \frac{{12\,\alpha _1^4\,c_6^a{\gamma ^2}}}{{h_0^3}} + \frac{{4\,c'h_0^3}}{{3\,\alpha _1^2{\gamma ^4}}} - \frac{{12\,c_6^b{\gamma ^2}{{\left( {2\,\alpha _1^2 + \alpha _3^2} \right)}^2}}}{{h_0^3}}\\
\left\langle  \frac{\partial^3 V}{\partial f_a \partial \phi_1^3 \partial \phi_3^1}\right\rangle  &=&
\frac{1}{{6\alpha _1^2{\alpha _3}{\gamma ^4}h_0^2}}\sqrt 2 {\mkern 1mu} \Big(36{\mkern 1mu} \alpha _1^3\alpha _3^3c_6^a{\gamma ^6} - 18{\mkern 1mu} \alpha _1^2\alpha _3^4c_6^a{\gamma ^6} - c'h_0^6 - 54{\mkern 1mu} \alpha _1^4\alpha _3^2c_6^a{\gamma ^6} + 144{\mkern 1mu} \alpha _1^3\alpha _3^3c_6^b{\gamma ^6}\nonumber\\
&&\quad \quad \quad \quad \quad \quad  + 72{\mkern 1mu} \alpha _1^5{\alpha _3}c_6^a{\gamma ^6} + 288{\mkern 1mu} \alpha _1^5{\alpha _3}c_6^b{\gamma ^6} + 24{\mkern 1mu} \alpha _1^3{\alpha _3}c_4^a{\gamma ^4}h_0^2 + 48{\mkern 1mu} \alpha _1^3{\alpha _3}c_4^b{\gamma ^4}h_0^2\nonumber\\
&&\quad \quad \quad \quad \quad \quad  - 12{\mkern 1mu} \alpha _1^2\alpha _3^2c_4^a{\gamma ^4}h_0^2\Big)
\\
\left\langle  \frac{\partial^3 V}{\partial f_b \partial \phi_1^3 \partial \phi_3^1}\right\rangle  &=&
\frac{1}{{3\,{\mkern 1mu} {\alpha _1}\alpha _3^2{\gamma ^4}h_0^2}}\Big(36\,{\mkern 1mu} \alpha _1^3\alpha _3^3c_6^a{\gamma ^6} - 54{\mkern 1mu} \alpha _1^2\alpha _3^4c_6^a{\gamma ^6} - c'h_0^6 - 18{\mkern 1mu} \alpha _1^4\alpha _3^2c_6^a{\gamma ^6} + 144{\mkern 1mu} \alpha _1^3\alpha _3^3c_6^b{\gamma ^6}\nonumber\\
&&\quad \quad \quad \quad \quad  + 72{\mkern 1mu} {\alpha _1}\alpha _3^5c_6^a{\gamma ^6} + 72{\mkern 1mu} {\alpha _1}\alpha _3^5c_6^b{\gamma ^6} + 24{\mkern 1mu} {\alpha _1}\alpha _3^3c_4^a{\gamma ^4}h_0^2 + 24{\mkern 1mu} {\alpha _1}\alpha _3^3c_4^b{\gamma ^4}h_0^2\nonumber\\
&&\quad \quad \quad \quad \quad  - 12{\mkern 1mu} \alpha _1^2\alpha _3^2c_4^a{\gamma ^4}h_0^2\Big)
\\
\left\langle  \frac{\partial^3 V}{\partial f_c \partial \phi_1^3 \partial \phi_3^1}\right\rangle  &=&
- \frac{1}{{3{\mkern 1mu} {\alpha _1}{\alpha _3}{\gamma ^4}h_0^3}}\Big(36{\mkern 1mu} \alpha _1^3\alpha _3^3c_6^a{\gamma ^6} - 36{\mkern 1mu} \alpha _1^2\alpha _3^4c_6^a{\gamma ^6} - 4{\mkern 1mu} c'h_0^6 - 36{\mkern 1mu} \alpha _1^4\alpha _3^2c_6^a{\gamma ^6} + 144{\mkern 1mu} \alpha _1^3\alpha _3^3c_6^b{\gamma ^6}\nonumber\\
&&\quad \quad \quad \quad \quad  + 36{\mkern 1mu} {\alpha _1}\alpha _3^5c_6^a{\gamma ^6} + 36{\mkern 1mu} \alpha _1^5{\alpha _3}c_6^a{\gamma ^6} + 36{\mkern 1mu} {\alpha _1}\alpha _3^5c_6^b{\gamma ^6} + 144{\mkern 1mu} \alpha _1^5{\alpha _3}c_6^b{\gamma ^6}\nonumber\\
&&\quad \quad \quad \quad \quad  - 12{\mkern 1mu} {\alpha _1}{\alpha _3}{c_2}{\gamma ^2}h_0^4\Big)
\\
\left\langle  \frac{\partial^3 V}{\partial \phi_3^1 \partial \phi_1^2 \partial s_2^3}\right\rangle  &=&
\frac{{18\,c_6^a\alpha _1^4\alpha _3^2{\gamma ^6} + 18\,c_6^a\alpha _1^2\alpha _3^4{\gamma ^6} + 12\,c_4^a\alpha _1^2\alpha _3^2{\gamma ^4}h_0^2 - c'h_0^6}}{{3\,\alpha _1^2{\alpha _3}{\gamma ^4}h_0^2}}\\
\left\langle  \frac{\partial^3 V}{\partial \eta_a \partial S_1^2 \partial \phi_2^1 }\right\rangle  &=&
\frac{{\sqrt 2 \Big( {24\,{c_3}h_0^6 - c'h_0^6 + 36\,\alpha _1^6c_6^a{\gamma ^6} + 12\,\alpha _1^4c_4^a{\gamma ^4}h_0^2} \Big)}}{{3\,\alpha _1^3{\gamma ^4}h_0^2}}\\
\left\langle  \frac{\partial^3 V}{\partial \eta_b \partial S_1^2 \partial \phi_2^1 }\right\rangle  &=&
\frac{{8\,{c_3}h_0^4}}{{\alpha _1^2{\alpha _3}{\gamma ^4}}}\\
\left\langle  \frac{\partial^3 V}{\partial f_a \partial \eta_a \partial \eta_a }\right\rangle  &=& 
\frac{{\sqrt 2 \,\Big( {48\,{c_3}h_0^6 - c'h_0^6 + 36\,\alpha _1^6c_6^a{\gamma ^6} + 144\,\alpha _1^6c_6^b{\gamma ^6} + 72\,\alpha _1^4\alpha _3^2c_6^b{\gamma ^6} + 12\,\alpha _1^4c_4^a{\gamma ^4}h_0^2 + 24\,\alpha _1^4c_4^b{\gamma ^4}h_0^2} \Big)}}{{3\,\alpha _1^3{\gamma ^4}h_0^2}}\nonumber\\
\\
\left\langle  \frac{\partial^3 V}{\partial f_a \partial \eta_a \partial \eta_b }\right\rangle  &=& 
\frac{{8\,{c_3}h_0^4}}{{\alpha _1^2{\alpha _3}{\gamma ^4}}}\\
\left\langle  \frac{\partial^3 V}{\partial f_a \partial \eta_b \partial \eta_b }\right\rangle  &=& 
\sqrt 2 \,\left( {8\,{\alpha _1}c_4^b + \frac{{24\,{\alpha _1}c_6^b{\gamma ^2}\left( {2\,\alpha _1^2 + \alpha _3^2} \right)}}{{h_0^2}}} \right)
\end{eqnarray}

\begin{eqnarray}
\left\langle  \frac{\partial^3 V}{\partial f_b \partial \eta_a \partial \eta_a }\right\rangle  &=& 
8\,{\alpha _3}c_4^b + \frac{{24\,{\alpha _3}c_6^b{\gamma ^2}\left( {2\,\alpha _1^2 + \alpha _3^2} \right)}}{{h_0^2}}\\
\left\langle  \frac{\partial^3 V}{\partial f_b \partial \eta_a \partial \eta_b }\right\rangle  &=& 
\frac{{8\sqrt 2 \,{c_3}h_0^4}}{{{\alpha _1}\alpha _3^2{\gamma ^4}}}\\
\left\langle  \frac{\partial^3 V}{\partial f_b \partial \eta_b \partial \eta_b }\right\rangle  &=& 
\frac{{2\Big( {24\,{c_3}h_0^6 - c'h_0^6 + 36\,\alpha _3^6c_6^a{\gamma ^6} + 36\,\alpha _3^6c_6^b{\gamma ^6} + 72\,\alpha _1^2\alpha _3^4c_6^b{\gamma ^6} + 12\,\alpha _3^4c_4^a{\gamma ^4}h_0^2 + 12\,\alpha _3^4c_4^b{\gamma ^4}h_0^2} \Big)}}{{3\,\alpha _3^3{\gamma ^4}h_0^2}}\nonumber\\
\\
\left\langle  \frac{\partial^3 V}{\partial f_c \partial \eta_a \partial \eta_a }\right\rangle  &=& 
\frac{{4\,{c_2}{h_0}}}{{{\gamma ^2}}} - \frac{{64\,{c_3}h_0^3}}{{\alpha _1^2{\gamma ^4}}} - \frac{{12\,\alpha _1^4c_6^a{\gamma ^2}}}{{h_0^3}} + \frac{{4\,c'h_0^3}}{{3\,\alpha _1^2{\gamma ^4}}} - \frac{{12\,c_6^b{\gamma ^2}{{\left( {2\alpha _1^2 + \alpha _3^2} \right)}^2}}}{{h_0^3}}\\
\left\langle  \frac{\partial^3 V}{\partial f_c \partial \eta_a \partial \eta_b }\right\rangle  &=& 
 - \frac{{32\sqrt 2 \,{c_3}h_0^3}}{{{\alpha _1}{\alpha _3}{\gamma ^4}}}\\
\left\langle  \frac{\partial^3 V}{\partial f_c \partial \eta_b \partial \eta_b }\right\rangle  &=& 
\frac{{4\,{c_2}{h_0}}}{{{\gamma ^2}}} - \frac{{32\,{c_3}h_0^3}}{{\alpha _3^2{\gamma ^4}}} - \frac{{12\,\alpha _3^4c_6^a{\gamma ^2}}}{{h_0^3}} + \frac{{4\,c'h_0^3}}{{3\,\alpha _3^2{\gamma ^4}}} - \frac{{12\,c_6^b{\gamma ^2}{{\left( {2\alpha _1^2 + \alpha _3^2} \right)}^2}}}{{h_0^3}}.
\end{eqnarray}

\end{document}